\documentclass[10pt]{article}
\usepackage[margin=0.7in]{geometry}
\usepackage{amsmath, amssymb, amsfonts, bbm, mathtools, enumitem, enotez, tikz, pgfplots, cancel, graphicx, caption, algorithm, algorithmic, hyperref}
\usetikzlibrary{matrix, arrows, positioning, angles}
\hypersetup{colorlinks=true, urlcolor=blue, linkcolor=violet}

\begin{document}\allowdisplaybreaks

\title{Spectral Derivatives}
\author{Pavel Komarov}
\date{January 8, 2025}
\maketitle

One of the happiest accidents in all math is the ease of transforming a function to and taking derivatives in the Fourier (i.e.~the \textit{frequency}) domain. But in order to exploit this extraordinary fact without serious artefacting, and in order to be able to use a computer, we need quite a bit of extra knowledge and care.

This document sets out the math behind the \href{https://pypi.org/project/spectral-derivatives/}{\texttt{spectral-derivatives}} Python package, all the way down to the bones, as much as I can manage. I try to get in to the real \textit{whys} behind what we're doing here, touching on fundamental signal processing and calculus concepts as necessary, and building upwards to more general cases.

\small
\tableofcontents
\normalsize

\section{Bases}

A \textit{basis} is a set of functions, call them $\{\xi_k\}$, that can be linearly combined to produce other functions. Often these are chosen to be orthogonal, meaning that if we take the ``inner product" of one funtion from the set with itself, we get back a constant (often normalized to be 1), and if we take the inner product of one of these functions with a different member of the set, we get back 0. In this sense the members of an orthogonal basis set are like perpendicular directions on a graph.

The inner product between two functions $f$ and $g$ is a generalization of the inner product between vectors, where instead of summing over a finite number of discrete entries, we integrate over infinitely many infinitesimally-separated points in the domain. We define it as:

$$ \langle f,g \rangle = \int\limits_{a}^{b} \overline{f(x)} g(x) dx $$

\noindent where the overbar $\overline{\circ}$ denotes a complex conjugate.

The inner product is symmetrical, so

$$ \langle f,g \rangle = \langle g,f \rangle = \int\limits_{a}^{b} f(x) \overline{g(x)} dx $$

\noindent Note that if we set $a$ and $b$ at $\pm \infty$, this integral could diverge. If it doesn't diverge with infinite bounds, we say the argument is \href{https://mathworld.wolfram.com/LebesgueIntegrable.html}{``Lebesgue integrable"}\cite{lebesgue}. Some of what we'll do only makes sense for this class of functions, so be aware.

\subsection{The Fourier Basis}\label{fourierB}

The most famous basis is the \textit{Fourier} basis,\endnote{\label{whyfourier}There's a great passage in \href{https://en.wikipedia.org/wiki/Richard_Hamming}{Richard Hamming}'s book \textit{The Art of Doing Science and Engineering}\cite{hamming} where he wonders why we use the Fourier basis so much:

\begin{quotation}
``It soon became clear to me digital filter theory was dominated by Fourier series, about which theoretically I had learned in college, and actually I had had a lot of further education during the signal processing I had done for John Tukey, who was a professor from Princeton, a genius, and a one or two day a week employee of Bell Telephone Laboratories. For about ten years I was his computing arm much of the time.

Being a mathematician I knew, as all of you do, that any complete set of functions will do about as good as any other set at representing arbitrary functions. Why, then, the exclusive use of the Fourier series? I asked various electrical engineers and got no satisfactory answers. One engineer said alternating currents were sinusoidal, hence we used sinusoids, to which I replied it made no sense to me. So much for the usual residual education of the typical electrical engineer after they have left school!

So I had to think of basics, just as I told you I had done when using an error-detecting computer. What is really going on? I suppose many of you know what we want is a time-invariant representation of signals, since there is usually no natural origin of time. Hence we are led to the trigonometric functions (the eigenfunctions of translation), in the form of both Fourier series and Fourier integrals, as the tool for representing things.

Second, linear systems, which is what we want at this stage, also have the same eigenfunctions—the complex exponentials which are equivalent to the real trigonometric functions. Hence a simple rule: if you have either a time-invariant system or a linear system, then you should use the complex exponentials.

On further digging in to the matter I found yet a third reason for using them in the field of digital filters. There is a theorem, often called Nyquist's sampling theorem (though it was known long before and even published by Whittaker, in a form you can hardly realize what it is saying, even when you know Nyquist's theorem), which says that if you have a band-limited signal and sample at equal spaces at a rate of at least two in the highest frequency, then the original signal can be reconstructed from the samples. Hence the sampling process loses no information when we replace the continuous signal with the equally spaced samples, provided the samples cover the whole real line. The sampling rate is often known as the Nyquist rate after Harry Nyquist, also of servo stability fame, as well as other things [also reputed to have been \href{https://deanebarker.net/tech/linkedin/harry-nyquist-again/}{just a really great guy} who often had productive lunches with his colleagues, giving them feedback and asking questions that brought out the best in them]. If you sample a non-band-limited function, then the higher frequencies are ``aliased" into lower ones, a word devised by Tukey to describe the fact that a single high frequency will appear later as a single low frequency in the Nyquist band. The same is not true for any other set of functions, say powers of $t$. Under equally spaced sampling and reconstruction a single high power of $t$ will go into a polynomial (many terms) of lower powers of $t$.

Thus there are three good reasons for the Fourier functions: (1) time invariance, (2) linearity, and (3) the reconstruction of the original function from the equally spaced samples is simple and easy to understand.

Therefore we are going to analyze the signals in terms of the Fourier functions, and I need not discuss with electrical engineers why we usually use the complex exponents as the frequencies instead of the real trigonometric functions. [It's down to convenience, really.] We have a linear operation, and when we put a signal (a stream of numbers) into the filter, then out comes another stream of numbers. It is natural, if not from your linear algebra course then from other things such as a course in differential equations, to ask what functions go in and come out exactly the same except for scale. Well, as noted above, they are the complex exponentials; they are the eigenfunctions of linear, time-invariant, equally spaced sampled systems.

Lo and behold, the famous transfer function [contains] exactly the eigenvalues of the corresponding eigenfunctions! Upon asking various electrical engineers what the transfer function was, no one has ever told me that! Yes, when pointed out to them that it is the same idea they have to agree, but the fact it is the same idea never seemed to have crossed their minds! The same, simple idea, in two or more different disguises in their minds, and they knew of no connection between them! Get down to the basics every time!"\end{quotation}

Back: \autoref{fourierB}} which is the set of complex exponentials:\vspace{-2mm}

\begin{equation}\label{euler}
e^{j \omega x} = \cos(\omega x) + j \sin(\omega x)
\end{equation}

\noindent where I use $j$ to represent the imaginary unit ($\sqrt{-1}$), because I'm from Electrical Engineering, and because Python uses \texttt{j}.

Why this identity is true isn't obvious at first but can be seen by \href{https://math.stackexchange.com/a/492165/278341}{Taylor Expanding}\cite{taylor} the exponential function and trigonometric functions:

$$e^a = 1 + a + \frac{a^2}{2!} + \frac{a^3}{3!} + ... = \sum_{n=0}^{\infty} \frac{a^n}{n!}$$

\noindent So\vspace{-2mm}

$$e^{j \omega x} = 1 + j \omega x + \frac{(j \omega x)^2}{2!} + \frac{(j \omega x)^3}{3!} + ... = 1 + j \omega x - \frac{(\omega x)^2}{2!} - j \frac{(\omega x)^3}{3!} + \frac{(\omega x)^4}{4!} - ...$$

$$\sin(\omega x) = \omega x - \frac{(\omega x)^3}{3!} + \frac{(\omega x)^5}{5!} - \frac{(\omega x)^7}{7!} + ...$$

$$\cos(\omega x) = 1 - \frac{(\omega x)^2}{2!} + \frac{(\omega x)^4}{4!} - \frac{(\omega x)^6}{6!} + ...$$\vspace{0mm}

\noindent Notice all the even-power terms appear with alternating sign as in the cosine expansion, and the odd-power terms appear with alternating sign as in the sine expansion, but with an extra $j$ multiplied in.

The presence of complex numbers to make this work can be confusing at first, but don't be scared! All we're really doing is using a compressed representation of a sine plus a cosine, where the real and imaginary parts (orthogonal in the complex plane, and therefore independent and non-interfering) allow us to describe the contributions of sine and cosine simultaneously. In fact, \href{https://math.stackexchange.com/a/1293127/278341}{Joseph Fourier originally used only real trigonometric functions}\cite{complex}, and it wasn't until later someone decided it would be easier to work with complex exponentials. Later (\autoref{fourierT}) we'll see that for real signals all the complex numbers cancel, leaving only a \textit{real} sine and \textit{real} cosine, which when added together make \textit{a single, phase-shifted sinusoid!} So think of $e^{j \omega x}$ as oscillations of frequency $\omega$ along direction $x$.

If we inner product sinusoidal wiggles of mismatched frequency, differing by $\Delta \omega$, they misalign and integrate to 0 over a period of length $T = \frac{2\pi}{\Delta \omega}$. But if we inner product matched wiggles, they align, multiply to 1 because of the inner product's complex conjugate, and integrate to a constant, the length of whichever domain we choose to consider, so this basis is orthogonal.

\section{Transforms}

We can construct a function from a linear combination of basis functions as:

\begin{equation}\label{spectralreconstruction}
f(x) = \sum_{k=0}^{\mathclap{M-1}} c_k \xi_k(x), \quad x \in [a, b]
\end{equation}

\noindent where $M$ is the number of basis functions we're using in our reconstruction and $k$ iterates through them. This is essentially a \textit{recipe}, which tells us \textit{how much} of each basis function $\xi_k$ is present in the signal $f$ on a domain between $a$ and $b$.

We can find the quantities $\{c_k\}$ by taking the inner product of the above with each of the basis functions to produce a system of $M$ equations, then solving. In the special case where $\{\xi_k\}$ are orthogonal, only one of the inner-producted terms from the sum will be nonzero, so the system's equations untangle to yield the simple relationship:

\begin{equation}\label{spectralcoefs}
c_k = \frac{\langle \xi_k,f \rangle}{\langle \xi_k,\xi_k \rangle} = \frac{\int\limits_{a}^{b} f(x) \overline{\xi_k(x)} dx}{||\xi_k||_2^2}
\end{equation}

This is completely analogous to a vector change-of-coordinates,\footnote{We are working with functions that live in $L^2$ space, L for Lebesgue, meaning functions which are square integrable on the domain, so the inner product integrals converge. Essentially, an element of this space is akin to a vector packed infinitely densely with values, until you get a function on a continuum. This is the only $L^p$ space which is also a Hilbert space, i.e.~has a nice inner product and ``induced norm" from that inner product which gives nice notions of distances analogous to standard Euclidean distance.} where $\vec{f}$ can be represented in terms of a new orthogonal basis $(\vec{\xi}_0, \vec{\xi}_1)$ instead of axis-aligned unit vectors $(\vec{e}_0,\vec{e}_1)$ via $\vec{f} = \frac{\langle \vec{\xi}_0,\vec{f}\rangle}{||\vec{\xi}_0||_2^2} \vec{\xi}_0 + \frac{\langle \vec{\xi}_1,\vec{f}\rangle}{||\vec{\xi}_1||_2^2} \vec{\xi}_1$:

\begin{center}
\begin{tikzpicture}[scale=3, >=stealth]
  \filldraw (0,0) circle (0.5pt);
  \draw[->, thick] (0,0) -- (1,0) node[anchor=west] {$\vec{e}_0$};
  \draw[->, thick] (0,0) -- (0,1) node[anchor=east] {$\vec{e}_1$};
  \draw[->, thick] (0,0) -- (0.9,1.2) node[anchor=west] {$\vec{f}$};
  \draw[->, thick, green!60!black] (0,0) -- (1.35, 0.45) node[anchor=south] {$\vec{\xi}_0$};
  \draw[->, thick, red] (0,0) -- (-0.2213594362, 0.66407830863) node[anchor=north east] {$\vec{\xi}_1$};
  \draw[dashed, gray] (0.9,1.2) -- (-0.27,0.81);
  \draw[dashed, gray] (0.9,1.2) -- (1.17, 0.39);
  \draw[->, dashed] (0,0) -- (1.17,0.39) node[below] {$\mathrm{proj}_{\vec{\xi}_0} \vec{f}$};
  \draw[->, dashed] (0,0) -- (-0.27,0.81) node[left] {$\mathrm{proj}_{\vec{\xi}_1} \vec{f}$};
\end{tikzpicture}
\end{center}

\noindent because:

\begin{align*}
\langle \vec{\xi}_k, \vec{f} - \mathrm{proj}_{\vec{\xi}_k} \vec{f} \rangle &= 0\quad\text{due to orthogonality}\\
\langle \vec{\xi}_k, \vec{f} - c_k\vec{\xi}_k \rangle &= 0\quad\text{because proj}_{\vec{\xi}_k}\vec{f} = c_k\vec{\xi}_k\\
\langle \vec{\xi}_k, \vec{f} \rangle - c_k \langle \vec{\xi}_k, \vec{\xi}_k \rangle &= 0\quad\text{by linearity}\\
\rightarrow c_k &= \frac{\langle \vec{\xi}_k, \vec{f} \rangle}{\langle \vec{\xi}_k, \vec{\xi}_k \rangle}
\end{align*}

\noindent The set of numbers $\{c_k\}$ is now \textit{an alternative representation} of the original function. In some sense it's equally descriptive, so long as we know which basis we're using to reconstruct. The function has been \textit{transformed}.

The $\{c_k\}$ are often said to live in another ``domain", although we have to be careful with this terminology, because it technically refers to a ``connected" set, not just a collection of $M$ things. To be precise, some authors instead use ``series" to describe $\{c_k\}$. However, it is possible for members of the basis set to be related through a continuous parameter which in some sense makes the set dense, even in cases where we only take discrete members of this more general set to be our basis set for a particular scenario. This is the case for the Fourier basis, where we choose $\omega \in \mathbb{R}$, and hence $\omega$ really can become a new domain.

\subsection{The Fourier Transform}\label{fourierT}

Using Fourier's original real-sinusoid-based formulation, we can write the reconstruction expression as\footnote{It's worth considering how weird it is this works to express arbitrary functions, even non-smooth ones (so long as they meet the Dirichlet conditions\cite{oppenheim}, i.e.~aren't pathological cases), a fact so counter-intuitive that Joseph Lagrange publicly declared Fourier was wrong at a meeting of the Paris Academy in 1807\cite{michigan} and rejected Fourier's paper, which then went unpublished until after Lagrange died!\cite{oppenheim} It's valuable to ask \href{https://math.stackexchange.com/questions/1105265/why-do-fourier-series-work}{why this works}\cite{why} and sift through \href{https://math.uchicago.edu/~may/REU2017/REUPapers/Xue.pdf}{some analysis}\cite{chicago}.}:

$$ f(x) = a_0 + \sum_{k=1}^{\infty} (a_k \cos(k \omega_0 x) + b_k \sin(k \omega_0 x))$$

where
\begin{itemize}[noitemsep, topsep=0pt, after=\newline]
	\item $f$ is periodic with fundamental frequency $\omega_0$, so the $k^\text{th}$ frequency becomes $k \cdot \omega_0$.
	\item $a_k$ and $b_k$ are coefficients describing how much cosine and sine to add in, respectively.
	\item $k$ goes up to $\infty$ because in general we need an infinite number of ever-higher-frequency sinusoids to reconstruct the function with perfect fidelity.
\end{itemize}

Let's now substitute $\cos(x) = \frac{e^{jx} + e^{-jx}}{2}$ and $\sin(x) = \frac{e^{jx} - e^{-jx}}{2j}$, which can be verified by manipulating Euler's formula, \autoref{euler}.\vspace{-2mm}

\begin{align*}
f(x) &= a_0 + \sum_{k=1}^{\infty} (a_k \frac{e^{j k \omega_0 x} + e^{-j k \omega_0 x}}{2} + b_k \frac{e^{j k \omega_0 x} - e^{-j k \omega_0 x}}{2j}) \\
&= a_0 + \sum_{\mathclap{k = -\infty}}^{-1} (\frac{a_{-k}}{2} - \frac{b_{-k}}{2j}) e^{j k \omega_0 x} + \sum_{k = 1}^{\infty} (\frac{a_k}{2} + \frac{b_k}{2j}) e^{j k \omega_0 x} = \sum_{\mathclap{k = -\infty}}^{\infty} c_k e^{j k \omega_0 x}
\end{align*}

\phantomsection\label{phase}
\noindent So if we choose $c_0 = a_0$ and $c_k = \overline{c_{-k}} = \frac{a_k}{2} + \frac{b_k}{2j}$, then the complex exponential formulation is \href{http://lpsa.swarthmore.edu/Fourier/Series/DerFS.html}{exactly equivalent to the trigonometric formulation}\cite{swarthmore}. That is, we can evaluate \autoref{spectralcoefs} using the \textit{complex} exponentials as basis functions to find \textit{complex} $c_k$ such that when multiplied in the reconstruction, \autoref{spectralreconstruction}, we get back only \textit{real} signal! Essentially, the relative balance of real and complex in $c_k$ affects how much cosine and sine are present at the $k^\text{th}$ frequency, \href{https://dsego.github.io/demystifying-fourier/}{thereby accomplishing a phase shift}\cite{sego}. Without accounting for phase shifts, we would only be able to model \textit{symmetric} signals!

If instead of a fundamental frequency $\omega_0 = \frac{2\pi}{T}$, where $T$ is a period of repetition, the signal contains dense frequencies (because it has no repetition, $T \rightarrow \infty$, $\omega_0 \rightarrow 0$), and if we care about a domain over the whole of $\mathbb{R}$, then it makes more sense to express both the basis functions and transformed coefficients as functions in $\omega$ and to make both our inner product and reconstruction expression integrals from $-\infty$ to $+\infty$:

\begin{equation}\label{pair}
\begin{aligned}
\text{inner product with }\xi\!:\quad\hat{f}(\omega) &= \int\limits_{-\infty}^{\infty} f(x) e^{-j \omega x} dx = \mathcal{F}\{f(x)\}\hspace{3cm} \\
\text{reconstruction}\!:\quad f(x) &= \frac{1}{2\pi} \int\limits_{-\infty}^{\infty} \hat{f}(\omega) e^{j \omega x} d \omega = \mathcal{F}^{-1}\{\hat{f}(\omega)\}
\end{aligned}
\end{equation}

\noindent where the hat $\hat{\circ}$ represents a function in the Fourier domain, and the $\frac{1}{2\pi}$ is a scaling factor that corrects for the fact $\hat{f}$ hasn't been normalized like the $\{c_k\}$ in the discrete case. Here, rather than divide by $||e^{j \omega x}||_2^2$, which diverges as the domain edges $\to \pm \infty$, we consider the domain $[\frac{-T}{2}, \frac{T}{2}]$ and let normalizing factor $T = \frac{2\pi}{\omega_0} \to \infty = \frac{2\pi}{d\omega}$\cite{oppenheim}.

Just like the $\{c_k\}$, $\hat{f}(\omega)$ can be complex, but if the original $f(x)$ is real, then $\hat{f}$'s complexity will perfectly interact with the complex exponentials to produce only a real function in the reconstruction.

\subsection{A Whole Family}\label{family}

Part of what makes Fourier transforms confusing is the proliferation of different variants for different situations, so it's worth \href{https://medium.com/sho-jp/fourier-transform-101-part-4-discrete-fourier-transform-8fc3fbb763f3 }{categorizing them}\cite{medium}. First off, are we dealing with a periodic signal (which has an $\omega_0$) or an aperiodic signal (which doesn't)? And second, are we dealing with a continuous function or discrete samples? Here is a diagram:

\begin{center}
\begin{tikzpicture}
\matrix (m) [
	cells={nodes={draw, minimum width=13em, minimum height=10em, outer sep=0pt}}
] at (0, 0) {
	\node (A) {}; & \node (B) {}; \\
	\node (C) {}; & \node (D) {}; \\
};

\node[anchor=south] at (A.north) {Periodic};
\node[anchor=south] at (B.north) {Aperiodic};
\node[anchor=east, rotate=90, yshift=2mm, xshift=10mm] at (A.west) {Continuous};
\node[anchor=east, rotate=90, yshift=2mm, xshift=7mm] at (C.west) {Discrete};

\node at (-3.25,2.5) (x_t_periodic) {$x(t)$}; 
\node at (-2,1.5) (X_ejw) {$X(e^{j\omega}$)};
\node at (3,3) (x_t_aperiodic) {$x(t)$}; 
\node at (3,1.5) (X_jw) {$X(j\omega$)};
\node at (-3, -1.5) (x_n_periodic) {$x[n]$};
\node at (-3, -3) (X_k) {$X[k]$};
\node at (2, -1.5) (x_n_aperiodic) {$x[n]$};
\node at (3.25, -2.5) (c_k) {$c_k$};

\draw[->, bend right=30] (X_ejw) to node[midway] {DTFT$^{-1}$} (x_n_aperiodic);
\draw[->, bend right=30] (x_n_aperiodic) to node[midway] {DTFT} (X_ejw);
\draw[->, bend left=45] (c_k) to node[midway] {FS$^{-1}$} (x_t_periodic);
\draw[->, bend left=45] (x_t_periodic) to node[midway] {FS} (c_k);
\draw[->, bend left=90] (X_k) to node[midway] {DFT$^{-1}$} (x_n_periodic);
\draw[->, bend left=90] (x_n_periodic) to node[midway] {DFT} (X_k);
\draw[->, bend left=90] (x_t_aperiodic) to node[midway] {FT} (X_jw);
\draw[->, bend left=90] (X_jw) to node[midway] {FT$^{-1}$} (x_t_aperiodic);
\end{tikzpicture}
\end{center}

\noindent Note that, following a more signal-processing-ish convention\cite{oppenheim}, the function we're transforming is now called $x$, and the independent variable, since it can no longer be $x$, is named $t$. For discrete signals, we use independent variable $n$ in square brackets.

Here FS stands for ``Fourier Series", which is the first situation covered above. FT stands for ``Fourier Transform", which is given by the integral pair, \autoref{pair}. But these are not the only possibilities! DTFT stands for ``Discrete Time Fourier Transform", where the signal we want to analyze is discrete but the transform is continuous. And finally DFT stands for ``Discrete Fourier Transform", not to be confused with the DTFT, which we use when \textit{both} the original and transformed signals are sampled.

\textit{All} of these can be considered Fourier transforms, but often when people talk about \textit{the} canonical ``Fourier Transform", they are referring to the continuous, aperiodic case in the upper righthand cell.

The notation of all these different functions and transforms is easy to mix up and made all the more confusing by the reuse of symbols. But it's important to keep straight which situation we're in. \href{https://www.youtube.com/watch?v=6ITWKtTYlEI&t=69s}{I can only apologize.} For more on all these, see \cite{oppenheim}.

\section{Taking Derivatives in the Fourier Domain}\label{derivative}

Let's \href{https://www.youtube.com/watch?v=d5d0ORQHNYs}{take a Fourier transform of the derivative of a function}\cite{brunton}:

$$\mathcal{F}\{\frac{d}{dx} f(x)\} = \int\limits_{-\infty}^{\infty} \underbrace{\frac{df}{dx}}_{dv} \underbrace{e^{-j \omega x}}_{u} dx = \underbrace{f(x) e^{-j \omega x} \Big|_{-\infty}^{\infty}}_{\mathclap{\parbox{30mm}{\footnotesize \centering 0 for Lebesgue-\\integrable functions}}} - \int\limits_{-\infty}^{\infty} f(x) (-j \omega) e^{-j \omega x} dx = j \omega \cdot \hat{f}(\omega)$$

\noindent We can use the inverse transform equation to see the same thing:

$$\frac{d}{dx} f(x) = \frac{d}{dx} \frac{1}{2\pi} \int\limits_{-\infty}^{\infty} \hat{f}(\omega) e^{j \omega x} d \omega = \frac{1}{2\pi} \int\limits_{-\infty}^{\infty} \hat{f}(\omega) \frac{d}{dx} e^{j \omega x} d \omega = \mathcal{F}^{-1}\{j \omega \cdot \hat{f}(\omega)\}$$

\noindent So a derivative in the $x$ domain can be accomplished by a \textit{multiplication} in the frequency domain. We can raise to higher derivatives simply by multiplying by $j \omega$ more times.

This is great because taking derivatives in the spatial domain is actually pretty hard, especially if we're working with discrete samples of a signal, whereas taking the derivative this way in the frequency domain, the \textit{spectral derivative}, gives us much better fidelity.\cite{trefethen4, kutz} The cost is that we have to do a Fourier transform and inverse Fourier transform to sandwich the actual differentiation, but there is an $O(N \log N)$ algorithm to accomplish the DFT (\autoref{family} and \autoref{dft}) for discrete signals called the Cooley-Tukey algorithm, also known as the Fast Fourier Transform (FFT)\cite{kutz}.

\subsection{Taking Derivatives in the Discrete Case}

Because we're going to want to use a computer, and a computer can only operate on discrete representations, we really need to talk about the DFT and what it means to take a derivative in this discrete paradigm. It has a connection to the above continuous case but is far more subtle, worth going in to \textit{\href{https://www.youtube.com/watch?v=v7l3Q11DBq0&t=1299s}{at some length}}.

\subsubsection{The DFT Pair}

\begin{equation}\label{dft}
\begin{aligned}
\text{DFT: \ \ } Y_k &= \sum_{n=0}^{\mathclap{M-1}} y_n e^{-j \frac{2\pi}{M} n k} \\
\text{DFT}^{-1} \text{: \ \ } y_n &= \frac{1}{M} \sum_{k=0}^{\mathclap{M-1}} Y_k e^{j \frac{2\pi}{M} n k}
\end{aligned}
\end{equation}

where
\begin{itemize}[noitemsep, topsep=0pt, after=\newline]
	\item $n$ iterates samples in the original domain (often spatial)
	\item $k$ iterates samples in the frequency domain (wavenumbers)
	\item $M$ is the number of samples in the signal, often given as $N$ by \href{https://numpy.org/doc/2.1/reference/routines.fft.html}{other sources}\cite{numpy}, but I'll use $N$ for something else later and want to be consistent
	\item $y$ denotes the signal in its original domain
	\item $Y$ denotes the signal in the frequency domain
\end{itemize}

We can express this as the linear inverse problem:

$$\frac{1}{M}\begin{bmatrix}
  e^{j \frac{2\pi}{M} 0 0} & e^{j \frac{2\pi}{M} 0 1} & \cdots & e^{j \frac{2\pi}{M} 0 (M-1)}\\
  e^{j \frac{2\pi}{M} 1 0} & e^{j \frac{2\pi}{M} 1 1} & \cdots & e^{j \frac{2\pi}{M} 1 (M-1)} \\
  \vdots & \vdots & \ddots & \vdots\\
  e^{j \frac{2\pi}{M} (M-1) 0} & e^{j \frac{2\pi}{M} (M-1) 1} & \cdots & e^{j \frac{2\pi}{M} (M-1) (M-1)}
\end{bmatrix}
\begin{bmatrix} Y_0 \\ Y_1 \\ \vdots \\ Y_{M-1} \end{bmatrix}
= \begin{bmatrix} y_0 \\ y_1 \\ \vdots \\ y_{M-1} \end{bmatrix}$$

\noindent which, thanks the special structure of its matrix, can be solved by a divide-and-conquer strategy that recursively builds the solution to a larger problem out of smaller ones (FFT)\cite{kutz}.

For simplicity, we can collect $\frac{2\pi}{M}n$ as a single term, $\theta_n \in [0, 2\pi)$, or $\frac{2\pi}{M}k$ as a single term, $\omega_k$. We then get $y_n = y(\theta_n)$ and $Y_k = Y(\omega_k)$. This may help highlight the fact the \href{https://dsp.stackexchange.com/a/18931/40873}{original signal and transformed signal live on a domain which maps to the unit circle}\cite{bristow} (hence periodicity and aliasing) and are being sampled at equally-spaced angles/angular velocities.

\subsubsection{Interpolation}

I now quote \href{https://math.mit.edu/~stevenj/fft-deriv.pdf}{Steven Johnson}\cite{johnson}, with some of my own symbols and notation sprinkled in:

\begin{quotation}\label{bandlimited}
``In order to compute derivatives like $y'(\theta)$, we need to do more than express $y_n$. We need to use the DFT$^{-1}$ expression to define a continuous interpolation between the samples $y_n$---this is called \textit{trigonometric interpolation}---and then differentiate this interpolation. At first glance, interpolating seems very straightforward: one simply evaluates the DFT$^{-1}$ expression at non-integer $n \in \mathbb{R}$. This indeed defines \textit{an} interpolation, but it is not the \textit{only} interpolation, nor is it the \textit{best} interpolation for this purpose. The reason there is more than one interpolation is due to \textit{aliasing}: any term $e^{+j \theta_n k} Y_k$ in the DFT$^{-1}$ can be replaced by $e^{+j \theta_n (k + mM)} Y_k$ for any integer $m$ and still give the \textit{same} samples $y_n$, since $e^{j \frac{2\pi}{M} nmM} = e^{j2\pi nm} = 1$ for any integers $m$ and $n$. Essentially, adding the $mM$ term to $k$ means that the interpolated function $y(\theta)$ just oscillates $m$ extra times between the sample points, which has no effect on $y_n$ but has a huge effect on derivatives. To resolve this ambiguity, one imposes additional criteria---e.g.~a bandlimited spectrum and/or minimizing some derivative of the interpolated $y(\theta)$"
\end{quotation}

We can now posit a slightly more general formula for the underlying continuous, periodic (over interval length M) signal:\vspace{-2mm}

$$ y(\theta) = \frac{1}{M} \sum_{k=0}^{\mathclap{M-1}} Y_k e^{j \theta (k + m_k M)}, \quad m_k \in \mathbb{Z} $$

\begin{quotation}
``In order to uniquely determine the $m_k$, a useful criterion is that we wish to \textit{oscillate as little as possible} between the sample points $y_n$. One way to express this idea is to assume that $y(\theta)$ is \textit{bandlimited} to frequences $|k + m_k M| \leq \frac{M}{2}$. Another approach, that gives the same result ... is to \textit{minimize the mean-square slope}"\footnote{It's due to this ambiguity and constraint that spectral methods are only suitable for smooth functions!}
\end{quotation}\vspace{-7mm}

\begin{align*}
\frac{1}{2\pi} \int\limits_{0}^{2\pi} |y'(\theta)|^2 d\theta &= \frac{1}{2\pi} \int\limits_{0}^{2\pi} \Big|\frac{1}{M} \sum_{k=0}^{M-1} j(k + m_k M) Y_k e^{j \theta (k + m_k M)} \Big|^2 d\theta \\
&= \frac{1}{2\pi M^2} \int\limits_{0}^{2\pi} \Big( \sum_{k=0}^{\mathclap{M-1}} j(k + m_k M) Y_k e^{j \theta (k + m_k M)} \Big) \overline{\Big( \sum_{k=0}^{\mathclap{M-1}} j(k + m_k M) Y_k e^{j \theta (k + m_k M)} \Big)} d\theta \\
&= \frac{1}{2\pi M^2} \int\limits_{0}^{2\pi} \sum_{k=0}^{M-1} \sum_{k'=0}^{M-1} \Big( j(k + m_k M) Y_k e^{j \theta (k + m_k M)} \Big) \overline{\Big( j(k' + m_{k'} M) Y_{k'} e^{j \theta (k' + m_{k'} M)} \Big)} d\theta \\
&= \frac{1}{M^2} \sum_{k=0}^{M-1} \sum_{k'=0}^{M-1} (k + m_k M) \overline{(k' + m_{k'} M)} Y_k \overline{Y_{k'}} \underbrace{\frac{1}{2\pi} \int\limits_{0}^{2\pi} e^{j \theta (k + m_k M)} e^{-j \theta (k' + m_{k'} M)} d\theta}_{\mathclap{= \begin{cases} 0 & \text{if } k + m_k M \neq k' + m_{k'} M \\ & \iff k \neq k' \text{ for } 0 \leq k, k' < M \\ 1 & \text{if } k = k'\end{cases}}} \\
&= \frac{1}{M^2} \sum_{k=0}^{\mathclap{M-1}} |Y_k|^2 (k + m_k M)^2
\end{align*}

We now seek to minimize this by choosing $m_k$ for each $k$. Only the last term depends on $m_k$, so it's sufficient to minimize only this:\vspace{-5mm}

\begin{align*}
\underset{m_k}{\text{minimize}} \quad & (k + m_k M)^2 \\
\text{s.t.} \quad & 0 \leq k < M \\
	& m_k \in \mathbb{Z}
\end{align*}

This problem actually admits of good ol' calculus plus some common sense:

$$\frac{d}{dm_k} (k + m_k M)^2 = 2(k + m_k M)M = 0 \longrightarrow m_k^* = \frac{-k}{M} \in (-1, 0]$$\vspace{0mm}

where $^*$ denotes optimality. But we additionally need to choose $m_k \in \mathbb{Z}$. Let's plot it to see what's going on.

\begin{center}
\begin{tikzpicture}
	\begin{axis}[axis lines=middle,
		xlabel={$m_k$}, xlabel style={below},
		ylabel={cost},
		xmin=-1.1, xmax=0.1, xtick={-1, -0.5, 0},
		ymin=-1, ymax=4.25, yticklabels={},
		grid=both,
		width=8cm, height=6cm]
		\addplot[domain=-1.25:0.25, samples=100, thick] {(1 + 3*x)^2};
		\addplot[only marks, mark=+, mark size=4pt] coordinates {(-1/3,0)};
		\node at (axis cs:-1/3,0) [anchor=north] {$m_k^*$};
		\addplot[only marks, mark=*] coordinates {(-1,4)};
		\addplot[only marks, mark=*] coordinates {(0,1)};
		\node at (axis cs:-0.5,3.25) {feasible costs};
		\draw[->, bend right=30] (axis cs:-0.7, 3.25) to (axis cs:-0.95,4);
		\draw[->, bend right=30] (axis cs:-0.4, 3) to (axis cs:-0.05,1.05);
	\end{axis}
\end{tikzpicture}
\end{center}

As we change the values of $M$ and $k$, the parabola shifts around, getting taller for larger $M$ and shifting leftward as $k \rightarrow M$.

We can see that for $k \in [0, \frac{M}{2})$, the $m_k = 0$ solution is lower down the cost curve, and for $k \in (\frac{M}{2}, M)$, the $m_k = -1$ solution is more optimal. ``If $k = \frac{M}{2}$ (for even $M$), however, there is an ambiguity: either $m_k = 0$ or $-1$ gives the same value $(k + m_k M)^2 = (\frac{M}{2})^2$. For this $Y_{M/2}$ term (the ``Nyquist" term), we can arbitrarily split up the $Y_{M/2}$ term between $m = 0$ [$j\frac{M}{2}\theta$, positive frequency] and $m = -1$ [$j(\frac{M}{2} - M)\theta = -j\frac{M}{2}\theta$, negative frequency]:"

$$Y_{M/2}(ue^{j\frac{M}{2}\theta} + (1 - u)e^{-j\frac{M}{2}\theta})$$\vspace{-7mm}

where $u \in \mathbb{C}$ s.t. at sample points $\theta_n$ we get $Y_{M/2}(ue^{j\frac{M}{2}\frac{2\pi}{M}n} + (1-u)e^{-j\frac{M}{2}\frac{2\pi}{M}n}) = Y_{M/2}(u\overbrace{e^{j\pi n}}^{(-1)^n} + (1-u)\overbrace{e^{-j\pi n}}^{(-1)^n}) = Y_{M/2}(-1)^n$ ``and so recover the DFT$^{-1}$."

If we use the above in the mean-squared slope derivation instead of $Y_k e^{j \theta (k + m_k M)}$ and $Y_{k'} e^{j \theta (k' + m_{k'} M)}$, then the \href{https://math.stackexchange.com/a/5013632/278341}{integral portion becomes}:\vspace{-5mm}

\begin{align*}
& Y_{M/2}\overline{Y_{M/2}}\frac{1}{2\pi} \int\limits_{0}^{2\pi} (ue^{j\frac{M}{2}\theta} + (1 - u)e^{-j\frac{M}{2}\theta}) \overline{(ue^{j\frac{M}{2}\theta} + (1 - u)e^{-j\frac{M}{2}\theta})} d\theta \\
&= |Y_{M/2}|^2 \frac{1}{2\pi} \Big(u\overline{u} \int\limits_{0}^{2\pi} \underbrace{e^{j\frac{M}{2}\theta} e^{-j\frac{M}{2}\theta}}_{= 1} d\theta + u \overline{(1 - u)} \int\limits_{0}^{2\pi} \underbrace{e^{j\frac{M}{2}\theta} e^{j\frac{M}{2}\theta}}_{\text{periodic!}} d\theta + (1 - u)\overline{u} \int\limits_{0}^{2\pi} \underbrace{e^{-j\frac{M}{2}\theta} e^{-j\frac{M}{2}\theta}}_{\text{periodic!}} d\theta \\
& \quad \quad + (1 - u) \overline{(1 - u)} \int\limits_{0}^{2\pi} \underbrace{e^{-j\frac{M}{2}\theta} e^{j\frac{M}{2}\theta}}_{= 1} d\theta \Big) \\
&= |Y_{M/2}|^2 \frac{1}{2\pi} (|u|^2 2\pi + |1 - u|^2 2\pi) = |Y_{M/2}|^2 (|u|^2 + |1 - u|^2)
\end{align*}

\noindent because integrating something periodic over a multiple of its period yields its mean, which is 0 in this case.

We now know that the contribution to the mean-squared slope from the $\frac{M}{2}^\text{th}$ term $\propto |u|^2 + |1 - u|^2$. What's the optimal $u$?\vspace{-2mm}

$$\frac{d}{du} |u|^2 + |1 - u|^2 = 2u - 2(1-u) = 0 \longrightarrow u = \frac{1}{2}$$

So ``the $Y_{M/2}$ term should be \textit{equally split} between the frequencies $\pm\frac{M}{2}\theta$, giving a $\cos(\frac{M}{2}\theta)$ term." Note that if $M$ is odd, there is no troublesome $\frac{M}{2}$ term like this, but later we'll use the \href{https://docs.scipy.org/doc/scipy/reference/generated/scipy.fft.dct.html}{Discrete Cosine Transform}\cite{dct} type I (DCT-I), which \hyperref[fftdct]{is equivalent to the FFT with even $M$ and $Y_k = Y_{M-k}$}, so we \textit{do} have to worry about the Nyquist term.

Now if we put it all together we get ``the \textbf{unique ``minimal-oscillation" trigonometric interpolation} of order $M$":

\begin{equation}\label{interpolant}
y(\theta) = \frac{1}{M} \Big(Y_0 + \sum_{\mathclap{0 < k < \frac{M}{2}}} \big(Y_k e^{j k \theta} + Y_{M-k} e^{-j k \theta}\big) + Y_{M/2}\cos(\frac{M}{2}\theta) \Big)
\end{equation}

``As a useful side effect, this choice of trigonometric interpolation has the property that real-valued samples $y_n$ (for which $Y_0$ is real and $Y_{M-k} = \overline{Y_k}$) will result in a purely real-valued interpolation $y(\theta)$ for all $\theta$."

\subsubsection{Taking Derivatives of the Interpolant}

Now at last, with this interpolation between integer $n$ in hand, we can take a derivative w.r.t. the spatial variable:

$$\frac{d}{d\theta} y(\theta) = \frac{1}{M} \Big( \sum_{\mathclap{0 < k < \frac{M}{2}}} j k (Y_k e^{j k \theta} - Y_{M-k} e^{-j k \theta}) - \frac{M}{2} Y_{M/2} \sin(\frac{M}{2}\theta) \Big)$$

Evaluating at $\theta_n = \frac{2\pi}{M}n, n \in \mathbb{Z}$, we get:\vspace{-5mm}

\begin{align*}
y'_n &= \frac{1}{M} \Big( \sum_{\mathclap{0 < k < \frac{M}{2}}} j k (Y_k e^{j k \frac{2\pi}{M}n} - Y_{M-k} e^{-j k \frac{2\pi}{M}n}) - \overset{\text{\large 0}}{\cancel{\frac{M}{2} Y_{M/2} \sin(\pi n)}} \Big) = \frac{1}{M} \sum_{k = 0}^{\mathclap{M-1}} Y'_k e^{j\frac{2\pi}{M}kn} \\
& \text{where} \quad Y'_k = \begin{cases} j k \cdot Y_k & k < \frac{M}{2} \\ 0 & k = \frac{M}{2} \\ j(k - M) \cdot Y_k & k > \frac{M}{2} \leftarrow \text{comes from: } k_{new} = M - k_{old}, 0 < k_{old} < \frac{M}{2} \end{cases}\\ & \hspace{5cm} \rightarrow \frac{M}{2} < k_{new} < M; -jk_{old} \cdot Y_{M - k_{old}} \rightarrow -j(M - k_{new}) \cdot Y_{k_{new}}
\end{align*}

Easy! Now let's do the second derivative:

$$\frac{d^2}{d\theta^2} y(\theta) = \frac{1}{M} \Big( \sum_{\mathclap{0 < k < \frac{M}{2}}} (jk)^2 (Y_k e^{j k \frac{2\pi}{M}n} + Y_{M-k} e^{-j k \frac{2\pi}{M}n}) - \Big(\frac{M}{2}\Big)^2 Y_{M/2} \cos(\frac{M}{2}\theta) \Big)$$

And again evaluating at $\theta_n = \frac{2\pi}{M}n, n \in \mathbb{Z}$:\vspace{-5mm}

\begin{align*}
y''_n &= \frac{1}{M} \Big( \sum_{\mathclap{0 < k < \frac{M}{2}}} (jk)^2 (Y_k e^{j k \frac{2\pi}{M}n} + Y_{M-k} e^{-j k \frac{2\pi}{M}n}) - \Big(\frac{M}{2}\Big)^2 Y_{M/2} (-1)^n \Big) = \frac{1}{M} \sum_{k = 0}^{\mathclap{M-1}} Y''_k e^{j\frac{2\pi}{M}kn} \\
& \text{where} \quad Y''_k = \begin{cases} (j k)^2 \cdot Y_k & k < \frac{M}{2} \\ \Big(j\frac{M}{2}\Big)^2 \cdot Y_k & k = \frac{M}{2} \\ (j(k - M))^2 \cdot Y_k & k > \frac{M}{2} \end{cases} \quad \text{or equivalently} \quad Y''_k = \begin{cases} (j k)^2 \cdot Y_k & k \leq \frac{M}{2} \\ (j(k - M))^2 \cdot Y_k & k > \frac{M}{2} \end{cases}
\end{align*}

It's important to realize ``this [second derivative] procedure is \textit{not} equivalent to performing the spectral first-derivative procedure twice (unless $M$ is odd so that there is no $Y_{M/2}$ term) because the first derivative operation omits the $Y_{M/2}$ term entirely."\cite{johnson}

We can repeat for higher derivatives, but the punchline is that for odd derivatives the $\frac{M}{2}$ term goes away,\footnote{For \textit{real} signals it is common (e.g.~in the kind of code ChatGPT might generate to do this) not to worry about zeroing out the Nyquist term and \textit{instead} throw away the imaginary part of the inverse transform. This works because $Y_{M/2} = \sum_{n=0}^{M-1} y_n e^{-j\frac{2\pi}{M} n \frac{M}{2}} = \sum_{n=0}^{M-1} y_n (-1)^n$, which will be purely real for real $y_n$, and when we mulitply by $(jk)^\nu$ for odd $\nu$, then $Y_{M/2}^{(\nu)}$ gets a constituent odd power of $j$, which makes it purely imaginary. Then its contribution to the inverse transform (i.e.~to each sample of the derivative $y_n^{(\nu)}$) is $+Y_{M/2}^{(\nu)} e^{j\frac{2\pi}{M}n\frac{M}{2}} = +Y_{M/2}^{(\nu)} (-1)^n$, which will also be purely imaginary. Whereas other imaginary components in the transform of a real signal have negated twins at negative frequency to pair with and become \textit{real} sines in the inverse transform, the Nyquist term has no twin, so it's the \textit{only} imaginary thing left over. Thus for \textit{real} signals keeping only the \texttt{ifft().real} is \textit{equivalent} to zeroing out the Nyquist term.} and for even derivatives it comes back. In general:

\begin{equation}\label{Y_nu}
Y^{(\nu)}_k = \begin{cases} (j k)^\nu \cdot Y_k & k < \frac{M}{2} \\ (j \frac{M}{2})^\nu \cdot Y_k & k = \frac{M}{2} \text{ and } \nu \text{ even} \\ 0 & k = \frac{M}{2} \text{ and } \nu \text{ odd} \\ (j(k - M))^\nu \cdot Y_k & k > \frac{M}{2} \end{cases}
\end{equation}

This has definite echoes of the standardly-given, continuous-time case covered in \autoref{derivative}, but it's emphatically \textit{not} as simple as just multiplying by $j\omega$ or even by $j k$. However, the final answer is thankfully super compact to represent in math and in code.

\subsection{Limitations}\label{artefacts}

So far it has all been good news, but there is a serious caveat to using the Fourier basis, especially for derivatives.

Although a Fourier transform tends to have more ``mass" at lower frequencies and fall off as we go to higher ones (otherwise the reconstruction integral would diverge), and therefore we can get really great reconstructions by leaving off higher modes\cite{kutz} (or equivalently only sampling and transforming $M$ components), we in fact need \textit{all} the infinite modes to reconstruct an \textit{arbitrary} signal\cite{oppenheim}. Even then, the Fourier basis can not represent true discontinuities nor non-smooth corners, instead converging ``almost everywhere", which is math speak for the ``measure" or volume of the set where it doesn't work being 0, meaning it only doesn't work \textit{at} the discontinuities or corners themselves.\cite{oppenheim}

If there are discontinuities or corners, we get what's called the Gibbs Phenomenon\cite{oppenheim} (see next figure), essentially overshoot as the set of basis functions tries to fit a sudden change. These extra wiggles are bad news for function approximation but even worse news for taking derivatives: If we end up on one of those oscillations, the slope might wildly disagree with that of the true function!

\begin{figure}[h!]
	\centering
	\includegraphics[width=0.7\textwidth]{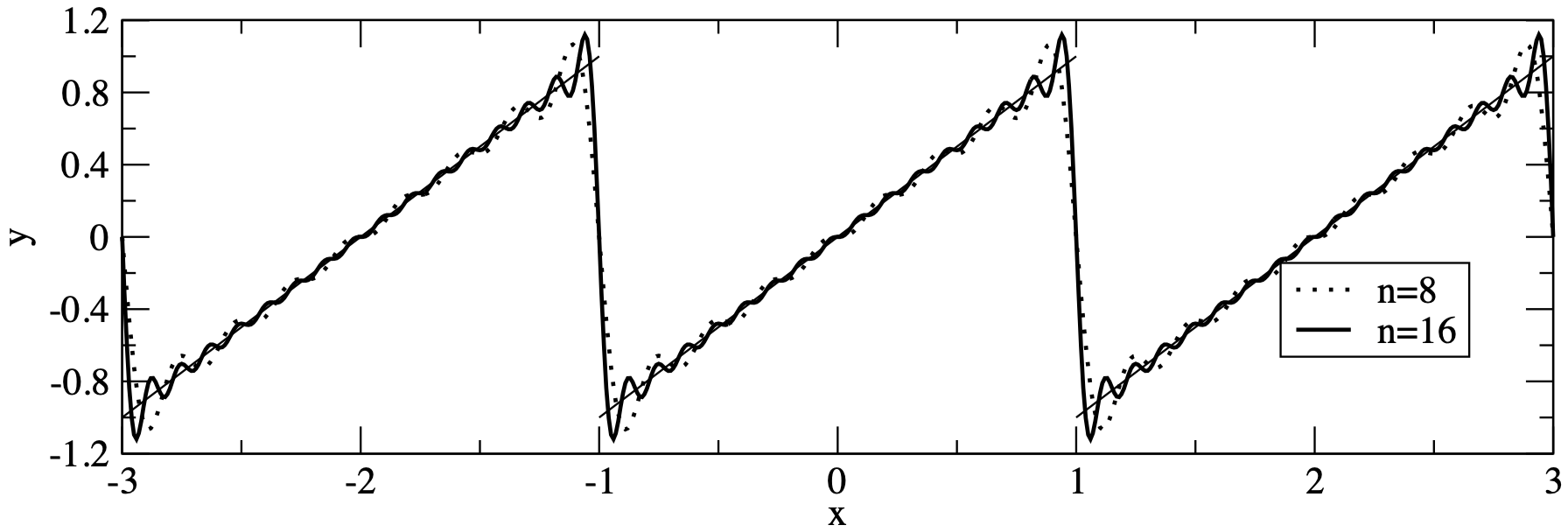}
	\caption*{\small An example of the Gibbs phenomenon, from \cite{kutz}. Using more modes allows the solution to fit better in the middle of the domain but makes peaks at the edges more extreme.}
\end{figure}

This is a bigger problem than it may first appear, because when we do this on a computer, we're using the DFT, which implicitly periodically extends the function (\autoref{family}, \autoref{dft}). So we not only need the function to have no jumps or corners internal to its domain; we need it to match up smoothly \textit{at the edges} of its domain too!

This rules out the above spectral method for all but ``periodic boundary conditions"\cite{kutz}. But if the story ended right there, I wouldn't have thought it worth building \href{https://pypi.org/project/spectral-derivatives/}{this package}.

\section{The Chebyshev Basis}

\href{https://www.youtube.com/watch?v=HloOBYPwlmU}{There is another} basis which we can use to represent arbitrary functions, called the \href{https://epubs.siam.org/doi/epdf/10.1137/1.9780898719598.ch8}{Chebyshev polynomials}\cite{trefethen8}, which has a really neat relationship to the Fourier basis.

\begin{figure}[h!]
\begin{center}
\begin{minipage}{0.45\textwidth}
\centering
\begin{tikzpicture}[scale=0.9]
	\begin{axis}[axis lines=middle, xlabel={Re$\{z\}$}, xlabel style={right, xshift=2mm},
		ylabel={Im$\{z\}$}, ylabel style={xshift=3mm, yshift=7mm}, xmin=-1.1, xmax=1.2, xtick={1},
		xticklabels={1}, xticklabel style={xshift=1mm, yshift=5mm}, ymin=-1.1, ymax=1.2, ytick=\empty,
		axis line style={->}, width=7.5cm, height=7.5cm]
		\addplot[domain=0:360,samples=100, smooth, thick] ({cos(x)}, {sin(x)}); 
		\pgfmathsetmacro{\angle}{50} 
		\pgfmathsetmacro{\xcoord}{cos(\angle)} 
		\pgfmathsetmacro{\ycoord}{sin(\angle)} 
		\addplot[only marks, mark=*, mark size=2pt] coordinates {(\xcoord, \ycoord)};
		\node at (axis cs:\xcoord, \ycoord) [above right] {$z = e^{j\theta}$};
		\addplot[only marks, mark=*, mark size=2pt] coordinates {(\xcoord, -\ycoord)};
		\node at (axis cs:\xcoord, -\ycoord) [below, right, yshift=-4mm] {\parbox{13mm}{\begin{align*}\overline{z} &= z^{-1} \\ &= e^{-j\theta}\end{align*}}};
		\addplot[domain=0:1, samples=2, thick] ({x*\xcoord}, {x*\ycoord}); 
		\addplot[domain=1:0, samples=2, thick, dashed] ({\xcoord}, {x*\ycoord}); 
		\addplot[only marks, mark=*, mark size=2pt] coordinates {(\xcoord, 0)};
		\node at (axis cs:\xcoord, 0) [above right] {$x$};
		\draw [thick](axis cs:0.1,0) ++(0:1) arc (0:\angle:0.3cm);
		\node at (axis cs:0.1, 0) [above right] {$\theta$};
	\end{axis}
\end{tikzpicture}
\end{minipage}
\begin{minipage}{0.45\textwidth}
	\begin{equation}\label{domains}
	\begin{aligned}
		\text{Let} \quad & x \in [-1, 1] \phantom{\frac{0}{0}} & \text{Chebyshev} \\ & = \cos(\theta),\ \theta \in [0, \pi] \phantom{\frac{0}{0}} & \text{Fourier} \\ & = \frac{1}{2}(z + z^{-1}),\ |z| = 1 & \text{Laurent}
	\end{aligned}
	\end{equation}
\end{minipage}
\end{center}
\end{figure}

The $k^\text{th}$ Chebyshev polynomial is defined as $T_k(x) = Re\{z^k\} = \frac{1}{2}(z^k + z^{-k}) = \cos(k\theta)$ by Euler formula:

\begin{align*}
T_0(x) &= Re\{z^0\} = 1\\
T_1(x) &= Re\{z^1\} = \frac{1}{2}(e^{j\theta} + e^{-j\theta}) = \cos(\theta) = x\\
T_2(x) &= \frac{1}{2}(e^{j2\theta} + e^{-j2\theta}) = \cos(2\theta)\\
& \text{but also } = \frac{1}{2}\underbrace{(z^2 + 2 + z^{-2})}_{\text{perfect square}} - 1 = \Big( \sqrt{\frac{1}{2}} (z + z^{-1}) \Big)^2 - 1 = \underbrace{\Big( \frac{2}{\sqrt{2}} \Big)^2}_{2} \underbrace{\frac{z + z^{-1}}{2}}_{\cos(\theta)} - 1 = 2x^2 - 1\\
T_3(x) &= \frac{1}{2}(e^{j3\theta} + e^{-j3\theta}) = \cos(3\theta)\\
& \text{but also } = \frac{1}{2}(z + z^{-1})^3 - \frac{3}{2}(z + z^{-1}) = 4x^3 - 3x\\
...
\end{align*}

It turns out there is a recurrent pattern:

$$ T_{k+1} = \frac{1}{2}(z^{k+1} + z^{-(k+1)}) = \frac{1}{2}(z^k + z^{-k})(z + z^{-1}) - \frac{1}{2}(z^{k-1} + z^{-(k-1)}) = 2xT_k(x) - T_{k-1}(x) $$

Due to the relationship between $\theta$ and $x$ on their respective domains, you can think of these polynomials as cosine waves ``wrapped around a cylinder and viewed from the side."\cite{trefethen8}

\begin{figure}[h]\label{cylinder}
	\centering
	\includegraphics[width=0.4\textwidth]{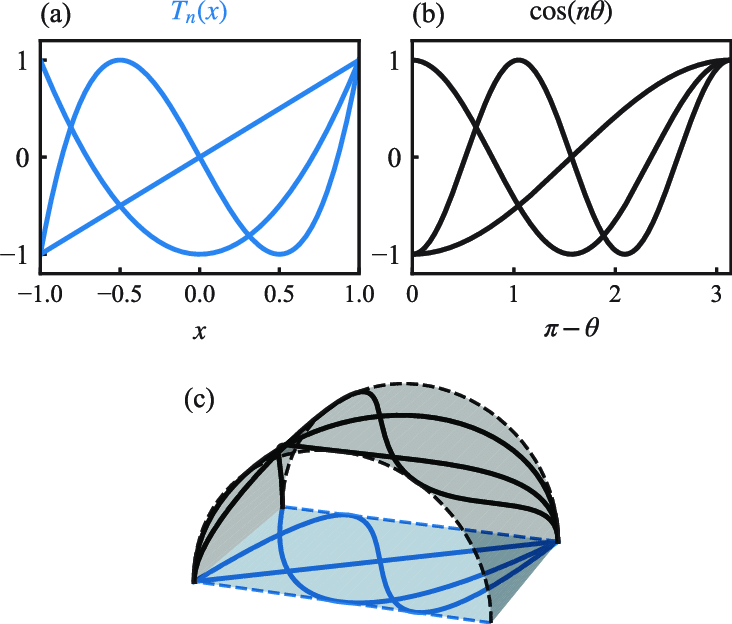}
	\captionsetup{width=0.75\textwidth}
	\caption*{\small Relationship of Chebyshev domain and Fourier Domain, from \cite{dedalus}. Notice the cosines are horizontally flipped. The authors use $n$ instead of $k$, which is common for Chebyshev polynomials (e.g. \cite{trefethen8}), but I prefer $k$ to enumerate basis modes, for consistency.}
\end{figure}

Essentially, on the domain $[-1, 1]$ each of these polynomials has ever more wiggles in the range $[-1, 1]$, and they perfectly coincide with the \textit{shadows} of horizontally-reversed $2\pi$-periodic cosines in the domain $[0, \pi]$. If we trace a function's value over $x = \cos(\theta)$ for linearly-increasing $\theta \in [0, \pi]$ instead of tracing it for linearly-decreasing $x \in [-1, 1]$, it's as if we're walking along the arc of the cylinder instead of along the shadow. We're effectively moving, horizontally flipping, and \textit{warping} the function (by expanding near the edges and compressing in the middle) to a new $\theta$ domain.

We can reconstruct a function using the different variables/basis formulations, and as long as our variables are related as in \autoref{domains}, these reconstructions are \textit{equivalent}:

\begin{equation}\label{equivalent}
y(x) = \sum_{k=0}^N a_k T_k(x)\ ;\quad y(z) = \sum_{k=0}^N a_k \frac{1}{2}(z^k + z^{-k})\ ;\quad y(\theta) = \sum_{k=0}^N a_k \cos(k \theta)
\end{equation}

Note the set of $\{a_k\}$ is for $k \in \{0, ... N\}$ and therefore has cardinality $N+1$.

\subsection{The Advantage of Chebyshev}

Why might we prefer this basis to the Fourier basis? Well, the advantage of a polynomial basis is we can avert the need for periodicity at the boundaries. Polynomial fits don't suffer the Gibbs Phenomenon, however they do suffer from the also-bad Runge Phenomenon\cite{kutz}:

\begin{figure}[h!]
	\centering
	\includegraphics[width=0.8\textwidth]{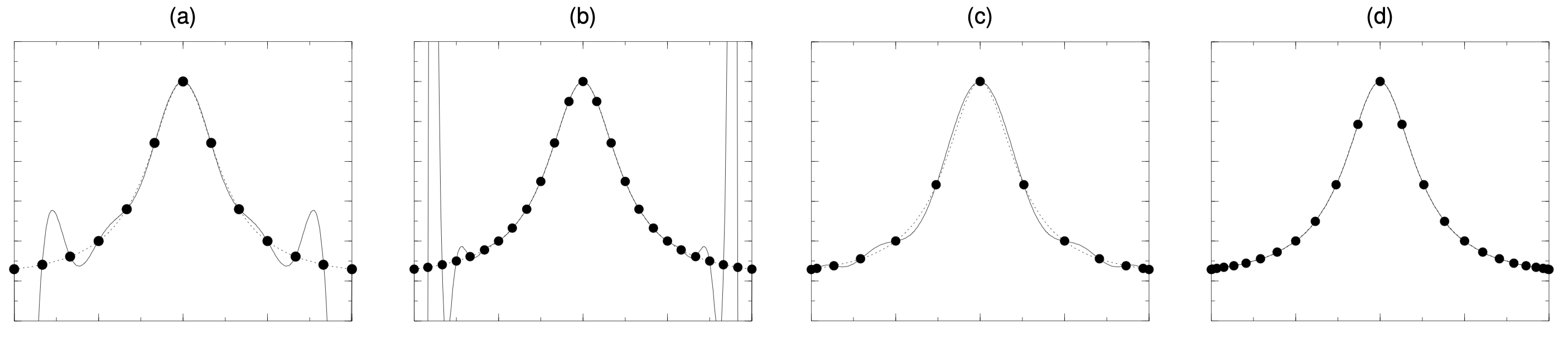}
	\captionsetup{width=\textwidth}
	\caption*{\small The Runge phenomenon, demonstrated in (a) and (b), mitigated in (c) and (d), from \cite{kutz}}
\end{figure}

However, there is something we can do about the Runge phenomenon: By clustering fit points at the edges of the domain, the wild wobbles go away.

If we take $x_n = \cos(\theta_n)$ with $\theta_n$ equispaced, $n \in \{0, ...N\}$, then we get a very natural clustering at the edges of $[-1, 1]$. What's more, if we have equispaced $\theta_n$ and a reconstruction expression built up out of sinusoids, we're back in a Fourier paradigm (at least in variable $\theta$) and can exploit the efficiency of the FFT, or, better, the discrete cosine and sine transforms!\cite{dct, dst}

Notice too that the polynomials/projected cosines are \textit{asymmetrical}, so we can natually use this basis to model arbitrary, lopsided functions without having to worry about \hyperref[phase]{phase shifts like we did for a Fourier basis of discrete harmonics}.

\section{Chebyshev Derivatives by Reduction to the Fourier Case}\label{algo}

This all suggests a solution procedure:

\begin{algorithm}
\caption*{\textbf{Chebyshev Derivative via Fourier}}
\begin{algorithmic}[1] 
\STATE Sample $y$ at $\{x_n = \cos(\theta_n)\}$ rather than at equally spaced $\{x_n\}$, thereby warping the function over the arc\\of a cylinder.
\STATE Use the DCT to transform to frequency domain.
\STATE Multiply by appropriate $(jk)^\nu$ to accomplish differentiation in the $\theta$ domain.
\STATE Inverse transform using the DST if odd function, DCT if even function.
\STATE Change variables back from $\theta$ to $x$, the Chebyshev variable, taking care that this entails an extra chain rule.
\end{algorithmic}
\end{algorithm}

\begin{figure}[h!]
\begin{center}
\begin{minipage}{0.45\textwidth}
\centering
\begin{tikzpicture}[scale=0.7]
	\begin{axis}[axis lines=middle, xlabel={$x$}, xlabel style={right},
		xmin=-1, xmax=1, xtick={-1, 0, 1},
		ymin=-2.5, ymax=2.5,
		width=8cm,
		height=5cm]
		\addplot[domain=-1:1, samples=100, thick] {exp(x)*sin(deg(7*x))};
		\foreach \n in {0, 1, ..., 10} {
			\addplot[only marks, mark=*, mark size=1.2pt] coordinates {(cos(deg(pi*\n/10)), {exp(cos(deg(pi*\n/10)))*sin(deg(7*cos(deg(pi*\n/10))))})};};
	\end{axis}
	\node at (3,3.75) {$e^x \sin(7x),\ x_n = \cos(\theta_n)$};
\end{tikzpicture}
\end{minipage}
\begin{minipage}{0.45\textwidth}
\centering
\begin{tikzpicture}[scale=0.8]
	\begin{axis}[axis lines=middle, xlabel={$\theta$}, xlabel style={right},
		xmin=0, xmax=pi, xtick={0, pi/2, pi}, xticklabels={0, $\frac{\pi}{2}$, $\pi$},
		ymin=-2.5, ymax=2.5,
		width=8cm,
		height=5cm]
		\addplot[domain=0:pi, samples=100, thick] {exp(cos(deg(x)))*sin(deg(7*cos(deg(x))))};
		\foreach \n in {0, 1, ..., 10} {
			\addplot[only marks, mark=*, mark size=1.2pt] coordinates {(pi*\n/10, {exp(cos(deg(pi*\n/10)))*sin(deg(7*cos(deg(pi*\n/10))))})};};
	\end{axis}
	\node at (3.25,3.75) {$e^{\cos(\theta)} \sin(7\cos(\theta)),\ \theta_n = \frac{2\pi}{10}n$};
\end{tikzpicture}
\end{minipage}
\begin{tikzpicture}[scale=0.8]
	\begin{axis}[axis lines=middle, xlabel={$\theta$}, xlabel style={right},
		xmin=-pi, xmax=2*pi, xtick={-pi, -pi/2, 0, pi/2, pi, 3*pi/2, 2*pi}, xticklabels={$-\pi$, $-\frac{\pi}{2}$, 0, $\frac{\pi}{2}$, $\pi$, $\frac{3\pi}{2}$, $2\pi$},
		ymin=-2.5, ymax=2.5,
		width=16cm,
		height=5cm]
		\addplot[domain=-pi:2*pi, samples=200, thick] {exp(cos(deg(x)))*sin(deg(7*cos(deg(x))))};
	\end{axis}
	\node at (9.5,0) {$e^{\cos(\theta)} \sin(7\cos(\theta)),\ \theta \in [-\pi, 2\pi]$};
\end{tikzpicture}
\begin{tikzpicture}[overlay, remember picture, scale=0.8]
	\draw[->, thick] (-8.7, 5.5) -- (-6.5, 5.5) node[midway, above] {\parbox{12mm}{\linespread{0.85}\selectfont flip and warp}};
	\draw[->, thick] (-4, 4) -- (-5, 3) node[midway, right, xshift=2mm] {\parbox{12mm}{\linespread{0.85}\selectfont periodically extend}};
\end{tikzpicture}
\end{center}
\captionsetup{width=0.85\textwidth}
\caption*{\small Illustration of implicit function manipulations in the first two steps of the algorithm. The edges of the aperiodic function can be made to match by ``periodic extension", but this operation alone only fixes discontinuity; corners are created at 0 and $\pi$, resulting in Gibbs phenomenon when we frequency transform. Warping \textit{stretches} corners into smooth transitions.}
\end{figure}

There are still a lot of details left to be worked out here, which we'll tackle in sequence.

\subsection{The Discrete Cosine Transform}\label{fftdct}

Because the reconstruction of $y(\theta)$ (\autoref{equivalent}) only contains cosines, doing a full FFT/DFT, which tries to fit sines as well as cosines, would be doing extra work. Instead we can use the DCT.

There are actually \href{https://docs.scipy.org/doc/scipy/reference/generated/scipy.fft.dct.html}{several possible definitions of the DCT}\cite{dct}, based on \href{https://en.wikipedia.org/wiki/Discrete_cosine_transform#Informal_overview}{different discrete periodic extensions}. The DCT-II is the default in \texttt{scipy}\cite{dct}, because it most closely resembles the \href{https://www.farbrausch.de/%7Efg/articles/dct_klt.pdf}{Karhunen–Loève Transform}\cite{klt}, which is provably optimal to represent signals generated by a stationary (statistical properties not varying over time) Markov process (where values are related only to \textit{immediate} previous values, through an autocorrelation coefficient). This class of signals is not exactly the same as the class of smooth functions, but it's similar in spirit. Thus we can often get better ``energy compaction" with this basis and \href{https://dsp.stackexchange.com/a/96197/40873}{represent signals in fewer coefficients}\cite{energycompaction}, especially when they have steep slopes at their boundaries. Indeed, common compression standards like JPEG choose to use the DCT-II.

However, we are dealing with \textit{warped} functions with \textit{flattened} edges, which are best made periodic by direct stacking-and-mirroring around the ends of the domain; they do not need any \href{https://github.com/pavelkomarov/spectral-derivatives/blob/main/notebooks/dct_types_comparison_and_derivatives.ipynb}{``wiggle room"} between repeats of the endpoints. To achieve this with the DCT-II, we have to sample at ``half-index" points, $\theta_{n,\text{II}} = \frac{1}{N+1}(n + \frac{1}{2})$, which do not go all the way to the the domain boundaries. But for compatibility with boundary value problems, we really want a sampling that includes the boundaries, like $\theta_n = \frac{\pi n}{N}$. The DCT-I implies the periodic extension we want with the sampling we want, so it is actually our best choice here, and it mercifully comes with the added benefit of being the least confusing variant to derive from the DFT:

Say we periodically extend $y$, i.e.~stack $y(\theta),\ \theta \in [0,\pi]$ next to a horizontal flip of itself on $(\pi, 2\pi)$, and then sample at the canonical DFT points, $\theta_n = \frac{2\pi}{M}n,\ n \in \{0, ... M-1\}$ (\autoref{dft}). We get:

$$\vec{y}_{\text{ext}} = \underbrace{[\underbrace{y_0, y_1, ... y_{N-1}, y_N}_{\parbox{20mm}{\footnotesize \centering original vector, length N+1}}, \underbrace{y_{N-1}, ... y_1}_{\parbox{16mm}{\footnotesize \centering redundant information}}]}_{\text{length M = 2N, necessarily even!}}, \text{ that is: } y_n = y_{M-n},\ 0 \leq n \leq N$$\vspace{2mm}

Then using $M-k$ for $k$ in the DFT equation, we get:

\begin{align*}
Y_{M-k} &= \sum_{n=0}^{M-1} y_n e^{-j \frac{2\pi}{M}n(M-k)} = \sum_{n=0}^{M-1} y_n \underbrace{e^{-j 2\pi n}}_{1} e^{j \frac{2\pi}{M}nk} = \sum_{n = M}^{1} y_{M-n} e^{j \frac{2\pi}{M}(M - n)k} \\
&= \sum_{n=1}^{M} \underbrace{y_{M-n}}_{= y_n} \underbrace{e^{j 2\pi k}}_{1} e^{-j\frac{2\pi}{M}nk} = \sum_{n=1}^M y_n e^{-j \frac{2\pi}{M}nk} = \sum_{n=0}^{M-1} y_n e^{-j \frac{2\pi}{M}nk} = Y_k \quad \square
\end{align*}
\begin{tikzpicture}[overlay, remember picture]
	\node at (11.5,3.25) {\footnotesize let $n_{new} = M - n_{old}$};
	\node at (11,0) {\parbox{27mm}{\footnotesize because $e^{-j\frac{2\pi}{M}Mk} = e^0 = 1$ and $y_M = y_0$}};
	\draw[->] (11.45,3.05) -- (11.37,2.5);
	\draw[->] (10.7,0.35) -- (10.86,0.95);
\end{tikzpicture}\newline

So when $y_n$ are redundant this way, the $Y_k$ are too, in a very mirror way. We can now use the facts $Y_k = Y_{M-k}$ and $N = \frac{M}{2}$ in the FFT interpolation (\autoref{interpolant}):

\begin{equation}\label{reconstruction}
\begin{aligned}
y(\theta) &= \frac{1}{M} \Big(Y_0 + \sum_{\mathclap{0 < k < \frac{M}{2}}} \big( Y_k e^{j k \theta} + Y_{M-k} e^{-j k \theta}\big) + Y_{M/2}\cos(\frac{M}{2}\theta) \Big) \\
&= \frac{1}{M} \Big(Y_0 + 2 \sum_{k = 1}^{N-1} \big( Y_k \underbrace{\frac{e^{j k \theta} + e^{-j k \theta}}{2}}_{\cos(k\theta)} \big) + Y_N\cos(N \theta) \Big)
\end{aligned}
\end{equation}

At samples $\theta_n = \frac{2\pi}{M}n = \frac{\pi}{N}n$, this becomes:

\[
y(\theta_n) = \frac{1}{M} \Big(Y_0 + Y_N\underbrace{\cos(\pi n)}_{(-1)^n} + 2 \sum_{k = 1}^{N-1} Y_k \cos(\frac{\pi nk}{N}) \Big) \tag{DCT-I$^{-1}$}
\]

This is exactly the DCT-I$^{-1}$, which, except for the $\frac{1}{M}$ term and a flip of $Y$ and $y$, is the same as the forward DCT-I! But the DCT and DCT$^{-1}$ operate on the shorter set of $\vec{Y} = [Y_0, ... Y_N]$, without redundant information. Thus

\begin{equation}\label{ifftidct}
\text{FFT}^{-1}(\underbrace{[Y_0, ... Y_N, Y_{N-1}, ... Y_1]}_{\vec{Y}_{\text{ext}}})[:\!N+1] = \text{DCT-I}^{-1}([Y_0, ... Y_N])
\end{equation}

Where $[:\!N+1]$ truncates to only the first $N+1$ elements ($\{0, ... N\}$). Given the equality above, we can line up everything we now know in a diagram:

\begin{figure}[h!]
\begin{center}
\begin{minipage}{0.5\textwidth}
\centering
\begin{tikzpicture}
	\node (y_ext) at (0, 2.25) {$\vec{y}_{\text{ext}}$};
	\node (Y_ext) at (0, 0) {$\vec{Y}_{\text{ext}}$};
	\node (y) at (4, 2.25) {$\vec{y}$};
	\node (Y) at (4, 0) {$\vec{Y}$};

	\draw[->, bend right=25] (y_ext) to node[midway, left, xshift=1mm] {FFT} (Y_ext);
	\draw[->, bend right=25] (Y_ext) to node[midway, right, xshift=-1mm] {FFT$^{-1}$} (y_ext);
	\draw[->, bend right=25] (y) to node[midway, left, xshift=1mm] {DCT-I} (Y);
	\draw[->, bend right=25] (Y) to node[midway, right, xshift=-1mm] {DCT-I$^{-1}$} (y);
	\draw[->, bend left=18] (y_ext) to node[midway, above, yshift=-0.5mm] {truncate} (y);
	\draw[->, bend left=18] (y) to node[midway, below, yshift=1.5mm] {truncate$^{-1}$} (y_ext);
	\draw[->, bend left=18] (Y_ext) to node[midway, above, yshift=-0.5mm] {truncate} (Y);
	\draw[->, bend left=18] (Y) to node[midway, below, yshift=1.5mm] {truncate$^{-1}$} (Y_ext);
\end{tikzpicture}
\end{minipage}
\begin{minipage}{0.3\textwidth}\raggedright where truncate$^{-1}$ does periodic extension, stacking back in redundant information\end{minipage}
\end{center}
\end{figure}

We can now easily see that in addition to the inverse relationship (\autoref{ifftidct}), we also have the forward relationship:

$$\text{FFT}([y_0, ... y_N, y_{N-1}, ... y_1])[:\!N+1] = \text{DCT-I}([y_0, ... y_N])$$

Notice that the $0^\text{th}$ and $N^\text{th}$ terms appear \textit{outside the sum}, and that the sum \textit{is multiplied by 2}. In our original conception of the cosine series for $y(\theta)$ (\autoref{equivalent}), all the cosines appear equally within the sum, so our $Y_k$ are subtly different from the $a_k$ in that formulation (some scaled by a factor of $\frac{1}{2}$ and all scaled by $M$). Both are valid, but it's more computationally convenient to use the DCT-based formulation.

\subsection{Even and Odd Derivatives and the Discrete Sine Transform}\label{evenodd}

The DCT can get us in to the frequency domain, but we'll need the help of another transform to get back out. We again start with the DCT-I formulation for simplicity.

If we look at the full $\vec{Y}_{\text{ext}}$ (\autoref{ifftidct}), we have a palindromic structure around $N$, but also around $0$, because of \href{https://dsp.stackexchange.com/a/18931/40873}{the repetitions}\cite{bristow}, which ensure we can read the values of $Y_k$ at negative $k$ by wrapping around to the end of the vector. This is describing an \textit{even} function, $f(-x) = f(x)$, which makes sense, because $y(\theta)$ is entirely composed of cosines, which are even functions, and because the forward transform is symmetrical with the inverse transform, the interpolation $Y(\omega)$ between $Y_k$ is also ultimately a bunch of cosines.

The derivative of an even function is an \textit{odd} function, $f(-x) = -f(x)$, which in principle should be constructable from purely sines, which are odd. And the derivative of an odd function is an even function again.

To see this more granularly, let's look in more detail at the multiplication by $(jk)^\nu$ that produces all the $Y_k^{(\nu)}$ (\autoref{Y_nu}), for $k \in \{0, ... M-1\}$:\vspace{-3mm}

\begin{align*}
\vec{Y}_{\text{ext}}^{(\nu)} &= [0, j^\nu, ... (j(N-1))^\nu, \underbrace{(\color{blue}0\color{black} \text{ or } \color{red}(jN)^\nu\color{black})}_{\parbox{20mm}{\footnotesize \centering depending on $\nu$ \color{blue}odd \color{black} or \color{red}even}}, (-j(N-1))^\nu, ... (-j)^\nu] \odot \vec{Y}_{\text{ext}}\\
&= j^\nu \cdot {\underbrace{[0, 1, ... 1, (\color{blue}0 \color{black}\text{ or }\color{red} 1\color{black}), -1, ..., -1]}_{\text{\large $\tilde{\mathbbm{1}}$}}}^\nu \odot \underbrace{[0, 1, ... N-1, N, N-1, ... 1]^\nu} \odot \vec{Y}_{\text{ext}}
\end{align*}
\begin{tikzpicture}[overlay, remember picture]
	\node at (4.35,0.25) {\footnotesize constant};
	\node at (13.4,0.2) {\footnotesize palindromic};
	\draw[->] (4.35,0.5) -- (4.35,1);
	\draw[->] (12.8,0.35) -- (11.95,0.8);
	\draw[->] (14,0.35) -- (14.6,1);
\end{tikzpicture}

where $\odot$ is a Hadamard, or element-wise, product, and raising a vector to a power is also element-wise. We can see

$$\tilde{\mathbbm{1}}^\nu = \begin{cases} [0, 1, ... 1, 0, -1, ..., -1] & \text{if $\nu$ is \color{blue}odd}\\ [0, 1, ... 1, 1, 1, ... 1] & \text{if $\nu$ is \color{red}even}\end{cases}$$

$[0, 1, ... 1, 0, -1, ..., -1]$ is odd around entries $0$ and $N$, and $[0, 1, ... 1, 1, 1, ... 1]$ is even around entry $0$.\newline

Let's now use this to reconstruct samples in the $\theta$ domain, $y_n^{(\nu)}$, for odd and even derivatives:\vspace{-5mm}

\begin{align}
y_n^{(\text{odd } \nu)} &= \frac{1}{M} \sum_{\mathclap{0 < k < \frac{M}{2}}} (jk)^\nu (Y_k e^{jk\theta_n} \color{blue}-\color{black} \underbrace{Y_{M-k}}_{= Y_k} e^{-jk\theta_n}) = \frac{1}{M} \sum_{k=1}^{N-1} (jk)^\nu Y_k \underbrace{(e^{jk\theta_n} - e^{-jk\theta_n})}_{2j\sin(k\theta_n)} \notag \\
&= \frac{1}{M} \underbrace{2 \sum_{k=1}^{N-1} (jk)^\nu Y_k j \sin(\frac{\pi nk}{N})}_{\text{\normalsize = a DST-I of $\vec{Y}^{(\nu)} \cdot j$!}} \label{y_prime_odd}\\
y_n^{(\text{even } \nu)} &= \frac{1}{M} \Big( \sum_{\mathclap{0 < k < \frac{M}{2}}} (jk)^\nu (Y_k e^{jk\theta_n} \color{red}+\color{black} \underbrace{Y_{M-k}}_{= Y_k} e^{-jk\theta_n}) + (j\frac{M}{2})^\nu Y_{M/2} \cos(\frac{M}{2}\theta_n) \Big) \notag \\
&= \frac{1}{M} \Big( (jN)^\nu Y_N \cos(\pi n) + \sum_{k=1}^{N-1} (jk)^\nu Y_k \underbrace{(e^{jk\theta_n} + e^{-jk\theta_n})}_{2\cos(k\theta_n)} \Big) \notag \\
&= \frac{1}{M} \Big( \underbrace{(j0)^\nu Y_0 + (jN)^\nu Y_N (-1)^n + 2 \sum_{k=1}^{N-1} (jk)^\nu Y_k \cos(\frac{\pi nk}{N})}_{\text{\normalsize = a DCT-I of $\vec{Y}^{(\nu)}$!}} \Big) \label{y_prime_even}
\end{align}
\begin{tikzpicture}[overlay, remember picture]
	\node at (7,7) {\parbox{13mm}{\footnotesize from oddness of $\tilde{\mathbbm{1}}^\nu$}};
	\node at (7.05,3.75) {\parbox{14mm}{\footnotesize from evenness of $\tilde{\mathbbm{1}}^\nu$}};
	\draw[->, blue] (7.5,7.3) -- (7.7,7.55);
	\draw[->, red] (7.65,4) -- (7.85,4.25);
\end{tikzpicture}

Brilliant! So we can use only the non-redundant $Y_k^{(\nu)}$ with a DST-I or DCT-I to convert odd and even functions, respectively, back to the $\theta$ domain!

Note that the DCT-I and DST-I definitions given in \texttt{scipy}\cite{dct, dst} use slightly different indexing than in my definitions here, which can be a point of confusion. I consistently take $N$ to be the \textit{index} of the last element of the non-redundant $y_n$, not its length, following \cite{trefethen8}. Note too that I consistently use $n$ to index samples and $k$ to index basis domain, whereas \texttt{scipy} uses $n$ for the domain being transformed \textit{from} and $k$ for the domain being transformed \textit{to}, which means these symbols are consistent with mine for forward transforms but flipped for inverse transforms.

Even more confusing, the DST-I only takes the $k \in \{1, ... N-1\}$ elements, since sines will result in zero crossings at $k = 0$ and $N$ (no informational content), whereas the DCT-I takes all $k \in \{0, ... N\}$ elements!

\subsection{Transforming Back to the Chebyshev Domain}

At this point we've accomplished all but the last step of \hyperref[algo]{the algorithm}, but we've been operating with $y_n = y(\theta_n)$ and $y_n^{(\nu)} = y^{(\nu)}(\theta_n)$, which are really samples from the $\theta$ domain, when what we really need to do is take derivatives in the $x = \cos(\theta)$ domain.

When we do this, we have to employ a chain rule, which introduces a new factor: the derivative of one of our variables w.r.t. the other. For the $1^\text{st}$ derivative it looks like:

$$\frac{d}{dx} y(\theta) = \frac{d}{d\theta} y(\theta) \cdot \frac{d\theta}{dx} = y'(\theta) \cdot \frac{d}{dx} \cos^{-1}(x) = y'(\theta) \cdot \frac{-1}{\sqrt{1 - x^2}}$$

The $y'(\theta)$ term is actually pretty easy to handle, because we know its value (and for higher orders too) at discretized $\theta_n$ from earlier! (Equations \ref{y_prime_odd} \& \ref{y_prime_even}) If we use the sampled $x_n = \cos(\theta_n)$ from step 1 of \hyperref[algo]{the algorithm}, then our $\{x_n\}$ and $\{\theta_n\}$ align, and we can find samples of the derivative w.r.t. $x$ by plugging $\{x_n\}$ in to the new factor(s) and multiplying appropriately (pointwise):

$$[\frac{d}{dx} y(\theta)]_n = \frac{-1}{\sqrt{1 - x_n^2}} \odot y'_n$$

\subsubsection{Higher Derivatives and Splintering Terms}

Let's see it for the second derivative:

\begin{align*}
\frac{d^2}{dx^2} y(\theta) &= \frac{d}{dx} \frac{y'(\theta)}{-\sqrt{1-x^2}} = \frac{-\sqrt{1-x^2} \frac{d}{dx} y'(\theta) - y'(\theta) \frac{d}{dx} (-\sqrt{1-x^2})}{1 - x^2} \\
&= \frac{-\frac{d}{d\theta}y'(\theta)\cdot \frac{d\theta}{dx}}{\sqrt{1-x^2}} - \frac{y'(\theta)\frac{x}{\sqrt{1-x^2}}}{1 - x^2} = \frac{y''(\theta)}{1-x^2} - \frac{x y'(\theta)}{(1 - x^2)^{3/2}} \\
\rightarrow [\frac{d^2}{dx^2} y(\theta)]_n &= \frac{1}{1-x_n^2} \odot y''_n - \frac{x_n}{(1 - x_n^2)^{3/2}} \odot y'_n
\end{align*}

Notice that the $2^\text{nd}$ derivative in $x$ requires \textit{both} the $1^\text{st}$ and $2^\text{nd}$ derivatives in $\theta$! This \textit{splintering} phenomenon will be a considerable source of pain as we take higher derivatives: For the $\nu^\text{th}$ derivative in $x$ we require all derivatives up to order $\nu$ in $\theta$.

Let's see a few more:

\begin{equation}\label{derivatives}
\footnotesize
\begin{aligned}
\frac{d^3}{dx^3} y(\theta) &= \frac{-1}{(1-x^2)^{3/2}} y'''(\theta) + \frac{3x}{(1-x^2)^2} y''(\theta) + \frac{-2x^2 - 1}{(1-x^2)^{5/2}} y'(\theta)\\
\frac{d^4}{dx^4} y(\theta) &= \frac{1}{(1-x^2)^2} y^{IV}(\theta) + \frac{-6x}{(1-x^2)^{5/2}} y'''(\theta) + \frac{11x^2 + 4}{(1-x^2)^3} y''(\theta) + \frac{-6x^3 - 9x}{(1-x^2)^{7/2}} y'(\theta)\\
\frac{d^5}{dx^5} y(\theta) &= \frac{-1}{(1-x^2)^{5/2}} y^V(\theta) + \frac{10x}{(1-x^2)^3} y^{IV}(\theta) + \frac{-35x^2-10}{(1-x^2)^{7/2}} y'''(\theta) + \frac{50x^2 + 55x}{(1-x^2)^4} y''(\theta) + \frac{-24x^4 -72x^2 - 9}{(1-x^2)^{9/2}} y'(\theta)\\
\end{aligned}
\normalsize
\end{equation}

We can see a bit of a pattern here, though. In particular, the function of $x$ multiplying each derivative of $y$ in $\theta$ comes from at most two terms in the preceding derivative, which have a predictable form:\vspace{-2mm}

\begin{align*}
\frac{d}{dx} \Big( \frac{p(x)}{(1-x^2)^{c-1}} y^{(\mu)}(\theta) &+ \frac{q(x)}{(1-x^2)^{c -\frac{1}{2}}} y^{(\mu-1)}(\theta) \Big) = \\
&\frac{p(x)}{(1-x^2)^{c-1}} y^{(\mu+1)}(\theta) \cdot \frac{d\theta}{dx} + \frac{(1 - x^2)^{c-1} \frac{d}{dx} p(x) - p(x)(c-1)(1-x^2)^{c-2}(-2x)}{(1-x^2)^{2c - 2}} y^{(\mu)}(\theta)\ +\\
&\frac{q(x)}{(1-x^2)^{c-\frac{1}{2}}} y^{(\mu)} \cdot \frac{d\theta}{dx} + \frac{(1 - x^2)^{c-\frac{1}{2}} \frac{d}{dx} q(x) - q(x)(c-\frac{1}{2})(1-x^2)^{c-\frac{3}{2}}(-2x)}{(1-x^2)^{2c - 1}} y^{(\mu-1)}(\theta)
\end{align*} 

\noindent If we now gather the $y^{(\mu)}(\theta)$ terms and use the fact $\frac{d\theta}{dx} = \frac{-1}{\sqrt{1-x^2}}$, we can find its new multiplying factor is equal to:\vspace{-2mm}

\begin{align*}
\frac{(1-x^2)p'(x) + 2(c-1)xp(x) - q(x)}{(1 - x^2)^c}
\end{align*}

This relationship holds no matter which, $\mu, c, p, q$ we're addressing, which allows us to build up a kind of pyramid of terms:

\begin{figure}[h!]\label{pyramid}
\begin{center}
\begin{minipage}{0.55\textwidth}
\centering
\begin{tikzpicture}
	\footnotesize
	\node[anchor=west] at (0, 4) {$-1$};
	\node[anchor=west] at (0, 3) {$-x$};\node[anchor=west] at (2.75, 3) {$1$};
	\node[anchor=west] at (0, 2) {$-2x^2-1$};\node[anchor=west] at (2.75, 2) {$3x$};\node[anchor=west] at (4.5, 2) {$-1$};
	\node[anchor=west] at (0, 1) {$-6x^3-9x$};\node[anchor=west] at (2.75, 1) {$11x^2+4$};\node[anchor=west] at (4.5, 1) {$-6x$};\node[anchor=west] at (6.5, 1) {$1$};
	\node[anchor=west] at (0, 0) {$-24x^4-72x^2-9$};\node[anchor=west] at (2.75, 0) {$50x^3+55x$};\node[anchor=west] at (4.5, 0) {$-35x^2-10$};\node[anchor=west] at (6.5, 0) {$10x$};\node[anchor=west] at (7.5, 0) {$-1$};
	\node at (0.5, -0.75) {lower $\frac{d}{d\theta}$ of $y$};
	\node at (8, -0.75) {higher $\frac{d}{d\theta}$ of $y$};
	\node at (-1, 4) {\parbox{15mm}{\linespread{0.85}\selectfont \normalsize numerator $p$ and $q$:}};
	\draw[->] (0,3.8) -- (0,0.2) node[midway, left] {\parbox{15mm}{\raggedleft increasing $\frac{d}{dx}$ of $y$}};
	\draw[->] (0.5,-0.55) -- (0.75,-0.2);
	\draw[->] (8,-0.55) -- (7.75,-0.2);
	\draw[->] (3,1.25) -- (3,1.75) node[midway, right] {p};
	\draw[->] (2.75,1) -- (1,1.75) node[midway, above] {q};
	\node at (3, -0.75) {...}; \node at (5, -0.75) {...};
\end{tikzpicture}
\end{minipage}
\begin{minipage}{0.4\textwidth}
\begin{tikzpicture}
	\node[anchor=west] at (0, 4) {$\frac{1}{2}$};
	\node[anchor=west] at (0, 3) {$\frac{3}{2}$};\node[anchor=west] at (1, 3) {$1$};
	\node[anchor=west] at (0, 2) {$\frac{5}{2}$};\node[anchor=west] at (1, 2) {$2$};\node[anchor=west] at (2, 2) {$\frac{3}{2}$};
	\node[anchor=west] at (0, 1) {$\frac{7}{2}$};\node[anchor=west] at (1, 1) {$3$};\node[anchor=west] at (2, 1) {$\frac{5}{2}$};\node[anchor=west] at (3, 1) {$2$};
	\node[anchor=west] at (0, 0) {$\frac{9}{2}$};\node[anchor=west] at (1, 0) {$4$};\node[anchor=west] at (2, 0) {$\frac{7}{2}$};\node[anchor=west] at (3, 0) {$3$};\node[anchor=west] at (4, 0) {$\frac{5}{2}$};
	\node at (1, -0.75) {...};\node at (4, -0.75) {...};
	\node at (-1, 4) {\parbox{15mm}{\linespread{0.85}\selectfont \normalsize $c$ at each location:}};
	\draw[->] (0,3.8) -- (0,0.2) node[midway, left] {\parbox{15mm}{\footnotesize \linespread{0.85}\selectfont \raggedleft increasing $\nu$}};
	\draw[->] (0.5,-0.5) -- (4.25,-0.5) node[midway, below, xshift=2mm] {\parbox{17mm}{\footnotesize increasing $\mu$}};
\end{tikzpicture}
\end{minipage}
\end{center}
\end{figure}

$q$ always refers to the element up and to the left, and $p$ always refers to the element above. If the arrows roll out of the pyramid, the corresponding $p$ or $q$ is $0$. I've done the above programmatically in \href{https://github.com/pavelkomarov/spectral-derivatives/blob/main/notebooks/alternative_chebyshev_derivative_methods.ipynb}{code}, such that we can find and apply the factors---and thereby accomplish the variable transformation back to the Chebyshev domain---for arbitrarily high derivatives.

\subsection{Handling Domain Endpoints}\label{endpoints}

There's a problem in the above at the edges of the domain: If $x = \pm 1$, the denominators of all the factors, which are powers of $\sqrt{1 - x^2} = 0$!

However, this doesn't mean $\frac{d^\nu}{dx^\nu} y$ can't have a valid \textit{limit value} at those points. First remember our reconstruction $y(\theta)$ is composed of cosines, so notice if we take odd derivatives in $\theta$, we get sines, and at the edges of the domain where $x \rightarrow \pm 1$, $\theta \rightarrow 0,\pi$, sine will be 0! However, if we take even derivatives, then $\cos(0,\pi) \rightarrow 1, -1$. Then, if we look closely at the derivatives in $x$ (\autoref{derivatives}), we can see that \textit{even} derivatives in $\theta$ of $y$ are divided by \textit{even} powers of $\sqrt{1 - x^2}$, and the \textit{highest} power in a denominator is an \textit{odd} power of $\sqrt{1 - x^2}$. If we multiply through so everything is over the highest-power denominator and then combine the expression into a single fraction, we get a situation where the odd-derivative terms are 0 because sines, and the even-derivative terms are 0 because they're multiplied by at least one $\sqrt{1 - x^2}$.

This means the \textit{numerator} as well as the denominator is $0$ at the domain endpoints. $\frac{0}{0}$ is an indeterminate form, so we can use L'Hôpital's rule!

Let's see it in fine detail for the $1^\text{st}$ derivative. The below uses the DCT-I reconstruction (\autoref{reconstruction}) for $y(\theta)$.

\begin{align*}
&\lim_{\substack{x \to \pm 1 \\ \theta \to 0,\pi}} \frac{d}{d\theta} \frac{1}{M} \Big( Y_0 + Y_N \cos(N\theta) + 2 \sum_{k=1}^{N-1} Y_k \cos(k\theta) \Big) \cdot \frac{d\theta}{dx} = \lim_{\substack{x \to \pm 1 \\ \theta \to 0,\pi}} \frac{\frac{1}{M} \Big(-N Y_N \sin(N\theta) - 2 \sum_{k=1}^{N-1} k Y_k \sin(k\theta) \Big)}{-\sqrt{1-x^2}} \\
&\underset{\frac{d}{dx}}{\overset{\frac{d}{dx}}{\longrightarrow}}\ = \lim_{\mathclap{\substack{x \to \pm 1 \\ \theta \to 0,\pi}}} \frac{\frac{1}{M} \Big(-N^2 Y_N \cos(N\theta) - 2 \sum_{k=1}^{N-1} k^2 Y_k \cos(k\theta) \Big) \cdot \frac{-1}{\cancel{\sqrt{1-x^2}}}}{\frac{x}{\cancel{\sqrt{1-x^2}}}} = \lim_{\mathclap{\substack{x \to \pm 1 \\ \theta \to 0,\pi}}} \frac{\frac{1}{M} \Big(N^2 Y_N \cos(N\theta) + 2 \sum_{k=1}^{N-1} k^2 Y_k \cos(k\theta) \Big)}{x}
\end{align*}

\[
= \begin{cases} \frac{1}{M} \Big(N^2 Y_N + 2 \sum_{k=1}^{N-1} k^2 Y_k \Big) & \text{ at } x=1, \theta=0 \\ -\frac{1}{M} \Big(N^2 (-1)^N Y_N + 2 \sum_{k=1}^{N-1} k^2 (-1)^k Y_k \Big) & \text{ at } x=-1, \theta=\pi \end{cases} \tag{$1^\text{st}$ endpoints}
\]

And now let's do it for the $2^\text{nd}$ derivative, with some slightly more compact notation, where we can be agnostic about $y(\theta)$'s exact structure until the end:

\begin{align*}
&\lim_{\substack{x \to \pm 1 \\ \theta \to 0,\pi}} \frac{\sqrt{1 - x^2} y''(\theta) - x y'(\theta)}{(1-x^2)^{3/2}} \rightarrow \frac{0}{0} \underset{\frac{d}{dx}}{\overset{\frac{d}{dx}}{\longrightarrow}} \frac{\cancel{\sqrt{1-x^2}}y'''(\theta)\frac{-1}{\cancel{\sqrt{1-x^2}}} + \cancel{\frac{-x}{\sqrt{1-x^2}}y''(\theta)} - (\cancel{xy''(\theta)\frac{-1}{\sqrt{1-x^2}}} + y'(\theta))}{-3x\sqrt{1-x^2}}\\
& \rightarrow \frac{0}{0} \underset{\frac{d}{dx}}{\overset{\frac{d}{dx}}{\longrightarrow}} \frac{\cancel{-}y^{IV}(\theta)\frac{\cancel{-1}}{\cancel{\sqrt{1-x^2}}} \cancel{-} y''(\theta)\frac{\cancel{-1}}{\cancel{\sqrt{1-x^2}}}}{\frac{6x^2 - 3}{\cancel{\sqrt{1-x^2}}}} = \frac{1}{6x^2 - 3}(y^{IV}(\theta) + y''(\theta))
\end{align*}

We already know $y''(\theta) = \frac{1}{M} \Big(-N^2 Y_N \cos(N\theta) - 2 \sum_{k=1}^{N-1} k^2 Y_k \cos(k\theta) \Big)$, assuming type I reconstruction. We can easily find

$$y^{IV}(\theta) = \frac{1}{M} \Big(N^4 Y_N \cos(N\theta) + 2 \sum_{k=1}^{N-1} k^4 Y_k \cos(k\theta) \Big)$$

Now we can evaluate these and the factor $\frac{1}{6x^2 - 3}$ at the limit values and put it all together to find: 

\[
\begin{cases} \frac{1}{3M} \Big((N^4 - N^2) Y_N + 2 \sum_{k=1}^{N-1} (k^4 - k^2) Y_k \Big) & \text{ at } x=1, \theta=0 \\ \frac{1}{3M} \Big((N^4 - N^2)(-1)^N Y_N + 2 \sum_{k=1}^{N-1} (k^4 - k^2) (-1)^k Y_k \Big) & \text{ at } x=-1, \theta=\pi \end{cases} \tag{$2^\text{nd}$ endpoints}
\]

\subsubsection{Endpoints for Higher Derivatives}

We can do the above for higher derivatives too. However, in general finding the endpoints for the $\nu^\text{th}$ derivative involves $\nu$ applications of L'Hôpital's rule, slowly cancelling one power of $\sqrt{1-x^2}$ at a time after each. The algebra gets to be pretty gnarly.

But there is some hope: We can see a pattern like the \hyperref[pyramid]{pyramid scheme} from earlier, because the functions multiplying each $y^{(\mu)}(\theta)$ in the numerator of the limit argument depend only on one or two terms from before the latest L'Hôpital. We can additionally use the relationship between variables $x = \cos(\theta)$ to recognize $\sqrt{1-x^2} = \sin(\theta)$ and substitute to put everything in terms of a single variable, and then just as well perform L'Hôpital's derivatives more simply w.r.t. $\theta$ rather than $x$ and cancel a $\sin(\theta)$ rather than a $\sqrt{1-x^2}$.

When we proceed, the denominator eventually acquires a single standalone term of the form $D\cos^\nu(\theta)$, which at the domain endpoints $0$ and $\pi$ will be something nonzero, $D_0 = \pm D_\pi$, thereby ending our \textit{journey}. At the same iteration, the numerator reduces to a set of constants, $C$, multiplying even-order $\theta$-derivatives of $y(\theta)$ up to the $2\nu^\text{th}$. Putting it all together, the endpoint formulas can be found as:

$$\begin{cases}
\frac{1}{D_0 M} \Big((... - C_3 N^6 + C_2 N^4 - C_1 N^2) Y_N + 2 \sum_{k=1}^{N-1} (... - C_3 k^6 + C_2 k^4 - C_1 k^2) Y_k \Big) & \text{ at } x=1, \theta=0 \\ \frac{1}{D_\pi M} \Big((... - C_3 N^6 + C_2 N^4 - C_1 N^2)(-1)^N Y_N + 2 \sum_{k=1}^{N-1} (... - C_3 k^6 + C_2 k^4 - C_1 k^2) (-1)^k Y_k \Big) & \text{ at } x=-1, \theta=\pi
\end{cases}$$

where the alternating plus and minus in the $k$ and $N$ terms comes from the fact the $2^\text{nd}$ derivative contains $-$cosines, the $4^\text{th}$ +cosines, the $6^\text{th}$ $-$cosines again, and so on.

Because the act of cancellation and the functions containing powers of $\sqrt{1-x^2} = \sin(\theta)$ can't be easily represented in \texttt{numpy}, computing $C$ and $D$ requires a symbolic solver like \texttt{sympy}. I've devised \href{https://github.com/pavelkomarov/spectral-derivatives/blob/main/notebooks/chebyshev_domain_endpoints.ipynb}{an implementation} to construct expressions for the endpoints, up to arbitrary order.

\section{Chebyshev Derivatives via Series Recurrence}

The above algorithm is a generalization of the method suggested by \href{https://epubs.siam.org/doi/epdf/10.1137/1.9780898719598.ch8}{Trefethen}\cite{trefethen8}, but it actually overcomplicates matters considerably. It is possible to sidestep all the ballooningly-difficult variable mappings and higher derivative limit-evaluations by exploiting \href{https://scicomp.stackexchange.com/questions/44939/chebyshev-series-derivative-in-terms-of-coefficients}{a direct relationship}\cite{chebder, brown, dcoefs} between a function's Chebyshev series and its derivative's Chebyshev series. This is analogous to how we can represent a function as a power series (a sum of integer powers of the independent variable) and find the coefficients of its derivative with \href{https://mathworld.wolfram.com/PowerRule.html}{Power Rule}\cite{powerrule}.

If we sandwich this key rule with the realization that the DCT is not only getting Fourier series coefficients of the warped function, but also (scaled) Chebyshev series coefficients of the original function (because these are exactly the same modulo constant factors!), and thus the DCT$^{-1}$ can go from the Chebyshev series representation to a cosine-spaced sampling of a function, then we have the ingredients to craft an algorithm that stays entirely in the $x$ domain:

\begin{algorithm}
\caption*{\textbf{Chebyshev Derivative via Series Rule}}
\begin{algorithmic}[1] 
\STATE Sample $y$ at $\{x_n = \cos(\theta_n)\}$ for equispaced $\{\theta_n\}$.
\STATE Use the DCT to get Chebyshev basis coefficients.
\STATE Use the Chebyshev series derivative rule to calculate the derivative's coefficients in $O(N)$.
\STATE Inverse transform with the DCT to resample the derivative function at $\{x_n\}$.
\end{algorithmic}
\end{algorithm}

This algorithm turns out to be \href{https://github.com/pavelkomarov/spectral-derivatives/blob/main/notebooks/alternative_chebyshev_derivative_methods.ipynb}{essentially numerically identical} to \hyperref[algo]{the first}, and due to its relative simplicity (no splinters!) and ease of extension to support non-cosine-spaced samples (albeit with considerably greater computational cost, $O(N^3)$ rather than $O(N \log N)$), it is the method of choice implemented in the main library \href{https://github.com/pavelkomarov/spectral-derivatives/blob/main/specderiv/specderiv.py}{code}.

\subsection{Chebyshev Series Derivative Rule}\label{seriesrule}

To most easily explain the rule, we'll need a little extra machinery: There are actually \textit{two different kinds} of Chebyshev polynomials, and so far we've only been working with the first kind, denoted $T_k(x) = \cos(k\theta)$. The \textit{second kind} is given as:

$$U_k(x) = \frac{\sin((k+1)\theta)}{\sin(\theta)}$$

\noindent Notice that:

$$\frac{d}{dx} T_k(x) = \frac{d}{d\theta} \cos(k\theta) \cdot \frac{d\theta}{dx} = k(\cancel{-}\sin(k\theta)) \cdot \frac{\cancel{-1}}{\sqrt{1-x^2}} = \frac{k\sin(k\theta)}{\sin(\theta)} = k \cdot U_{k-1}(x)$$

But we really want the derivative in terms of $T$, not $U$. Lucky for us, there is a relationship between the two, based on trigonometric identities, making particular use of $\cos(\alpha)\sin(\beta) = \frac{1}{2}(\sin(\alpha + \beta) - \sin(\alpha - \beta))$:

\begin{align*}
2\!\!\!\!\sum_{\text{odd }\kappa>0}^{\text{odd }k-1} \!\!\!\!&\cos(\kappa\theta)\sin(\theta) = \cancel{2}\!\!\!\!\sum_{\text{odd } \kappa>0}^{\text{odd }k-1} \cancel{\frac{1}{2}}(\sin((\kappa+1)\theta) - \sin((\kappa-1)\theta)\\
&= \sin(k\theta) \cancel{-\sin((k-2)\theta) + \sin((k-2)\theta)} - ... \cancel{-\sin(2\theta) + \sin(2\theta)} - \overset{\text{\normalsize 0}}{\cancel{\sin(0\theta)}} = \sin(k\theta)\\
&\rightarrow \frac{\sin(k\theta)}{\sin(\theta)} = 2\!\!\!\!\sum_{\text{odd }\kappa>0}^{k-1} \!\!\!\!\cos(\kappa\theta) \quad \text{for even }k
\end{align*}

\noindent Similarly:

\begin{align*}
2\!\!\!\!\sum_{\text{even }\kappa \geq 0}^{\text{even }k-1} \!\!\!\!&\cos(\kappa\theta)\sin(\theta) = \cancel{2}\!\!\!\!\sum_{\text{even } \kappa \geq 0}^{\text{even }k-1} \cancel{\frac{1}{2}}(\sin((\kappa+1)\theta) - \sin((\kappa-1)\theta)\\
&= \sin(k\theta) \cancel{-\sin((k-2)\theta) + \sin((k-2)\theta)} - ... \cancel{-\sin(\theta) + \sin(\theta)} \underbrace{-\sin(-\theta)}_{=+\sin(\theta)} = \sin(k\theta) + \sin(\theta)\\
&\rightarrow \frac{\sin(k\theta)}{\sin(\theta)} = -1 + 2\!\!\!\!\sum_{\text{odd }\kappa>0}^{k-1} \!\!\!\!\cos(\kappa\theta) \quad \text{for odd }k
\end{align*}

\noindent Thus:

$$k \cdot U_{k-1}(x) = \frac{d}{dx} T_k(x) = k \cdot \begin{cases}
2\sum_{\text{odd } \kappa>0}^{k-1} T_\kappa(x) & \text{for even } k \\
-1 + 2\sum_{\text{even } \kappa \geq 0}^{k-1} T_\kappa(x) & \text{for odd } k
\end{cases}$$

Let's see this on a couple examples to get a better intuition: In practice we represent a function with $N$ Chebyshev series coefficients, stored low to high. If $N=5$, then $T_3(x)$ would be $[0, 0, 0, 1, 0]$, and $T_4(x)$ would be $[0, 0, 0, 0, 1]$. If we differentiate these two, we should get $\frac{d}{dx}T_3(x) = 3 \cdot(-1 + 2 \cdot [1, 0, 1, 0, 0]) = [3, 0, 6, 0, 0]$, and $\frac{d}{dx}T_4(x) = 4 \cdot 2 \cdot [0, 1, 0, 1, 0] = [0, 8, 0, 8, 0]$. Notice the extra constant from the $-1$ in the first example is factored in to the coefficient of $T_0(x) = 1$.

Because differentiation is linear, we can scale and stack these particular results, e.g. $\frac{d}{dx}(2T_3(x) + T_4(x)) = 6T_0(x) + 8T_1(x) + 12T_2(x) + 8T_3(x)$. And because each term's derivative only affects every-other term of lower order, and these effects are cumulative, it's possible to calculate the new sequence by starting at the higher-order end and working downwards, modifying only \href{https://scicomp.stackexchange.com/q/44939/48402}{two numbers at each step}\cite{dcoefs}. An implementation of this procedure called \href{https://github.com/numpy/numpy/blob/v2.2.0/numpy/polynomial/chebyshev.py#L874-L961}{\texttt{chebder}}\cite{chebder} lives in \texttt{numpy}.

\section{Multidimensionality}

We are now fully equipped to find derivatives for 1-dimensional data. This is technically all we need, because, due to linearity of the derivative operator, we can find the derivative along a particular dimension of a multidimensional space by using our 1D solution along each constituent vector running in that direction, and we can find derivatives along multiple dimensions by applying the above in series along each dimension:

$$\frac{\partial^2}{\partial x_1 \partial x_2} y(x_1, x_2) = \texttt{Algo}(\texttt{Algo}(y_i, 1^\text{st}, x_1)_j, 1^\text{st}, x_2) \quad \forall\ i, j$$

$$ \nabla^2 y = (\frac{\partial^2}{\partial x_1^2} + \frac{\partial^2}{\partial x_2^2}) y = \texttt{Algo}(y_i, 2^\text{nd}, x_1) + \texttt{Algo}(y_j, 2^\text{nd}, x_2) \quad \forall\ i, j$$

\noindent where $i, j$ are indexers as in the computing sense and have nothing to do with the imaginary unit, \texttt{Algo} applies the algorithm to each vector along the dimension given by the third argument, and the $1^\text{st}$ and $2^\text{nd}$ in the second argument refer to the derivative order.

Each application to a vector incurs $O(N \log N)$ cost, and fundamentally applying the method to higher-dimensional data must involve a loop, so the full cost of applying along any given direction is (assuming length $N$ in all dimensions) $O(N^D \log N)$, where $D$ is the dimension of the data. Aside from pushing this loop lower down into \texttt{numpy} to take advantage of vectorized compute, there can be no cost savings for a derivative in a particular dimension.

\subsection{Dimensions Together versus In Series}

Can we simplify the situation at all?

Due to the linearity of the Fourier transform (and DCT), transforming along all dimensions, multiplying by appropriate $(jk)^\nu$ (or calling \texttt{chebder}) along corresponding dimensions of the transformed data, and then inverse transforming along all dimensions is equivalent to transforming, differentiating, and inverse transforming each dimension in series\cite{johnson}:

\begin{center}
\begin{tikzpicture}
	\node[draw, rectangle, minimum width=2.5cm, minimum height=2.5cm] (y) {$y$};
	\node[draw, rectangle, minimum width=2.5cm, minimum height=2.5cm, right=of y] (Y) {$Y$};
	\node[draw, rectangle, minimum width=2.5cm, minimum height=2.5cm, right=of Y] (jkY) {$(jk)^{\tilde{\nu}} \odot Y$};
	\node[draw, rectangle, minimum width=2.5cm, minimum height=2.5cm, right=of jkY] (dy) {$dy$};
	\draw[->] (y) -- (Y);
	\draw[->] (Y) -- (jkY);
	\draw[->] (jkY) -- (dy);
	\draw[->] (-1,1.5) -- (-1,-1.5) node[midway, above left, yshift=3mm, xshift=1mm] {\footnotesize FFT};
	\draw[->] (-1.5,-1) -- (1.5,-1) node[midway, above] {\footnotesize FFT};
	\draw[<->] (2.5,1.5) -- (2.5,-1.5) node[midway, above left, yshift=3mm, xshift=1mm] {\footnotesize $(jk)^{\nu_d}\odot$};
	\draw[<->] (2,-1) -- (5,-1) node[midway, above] {\footnotesize $\odot (jk)^{\nu_{d'}}$};
	\draw[<-] (6,1.5) -- (6,-1.5) node[midway, above left, yshift=3mm, xshift=1mm] {\footnotesize FFT$^{-1}$};
	\draw[<-] (5.5,-1) -- (8.5,-1) node[midway, above] {\footnotesize FFT$^{-1}$};
	\node[anchor=north, yshift=-1mm] at (y.south) {$O(D \cdot N^D \log N)$};
	\node[anchor=north, yshift=-1mm] at (Y.south) {$O(D \cdot N^D)$};
	\node[anchor=north, yshift=-1mm] at (jkY.south) {$O(D \cdot N^D \log N)$};
\end{tikzpicture}
\end{center}

\noindent That's neat, but does it save us anything, really? Let's see it in series:

\begin{center}
\begin{tikzpicture}
	\node[draw, rectangle, minimum width=2.5cm, minimum height=2.5cm] (y) {$y$};
	\node[draw, rectangle, minimum width=2.5cm, minimum height=2.5cm, right=of y] (Y1) {$Y_d$};
	\node[draw, rectangle, minimum width=2.5cm, minimum height=2.5cm, right=of Y] (jkY1) {$(jk)^{\nu_d} \odot Y_d$};
	\node[draw, rectangle, minimum width=2.5cm, minimum height=2.5cm, right=of jkY] (dy1) {$\partial_d y$};
	\node[draw, rectangle, minimum width=2.5cm, minimum height=2.5cm, below=of dy1] (Y2) {$Y_{d'}$};
	\node[draw, rectangle, minimum width=2.5cm, minimum height=2.5cm, left=of Y2] (jkY2) {$(jk)^{\nu_{d'}} \odot Y_{d'}$};
	\node[draw, rectangle, minimum width=2.5cm, minimum height=2.5cm, left=of jkY2] (dy2) {$\partial_{d'} \partial_{d} y$};
	\node[left=of dy2, font=\huge] (dots) {...};
	\draw[->] (y) -- (Y);
	\draw[->] (Y) -- (jkY);
	\draw[->] (jkY) -- (dy1);
	\draw[->] (dy1) -- (Y2);
	\draw[->] (Y2) -- (jkY2);
	\draw[->] (jkY2) -- (dy2);
	\draw[->] (dy2) -- (dots);
	\draw[->] (-1,1.5) -- (-1,-1.5) node[midway, above left, yshift=3mm, xshift=1mm] {\footnotesize FFT};
	\draw[->] (9,-1) -- (12,-1) node[midway, above] {\footnotesize FFT};
	\draw[<->] (9,-4.5) -- (12,-4.5) node[midway, above] {\footnotesize $\odot(jk)^{\nu_{d'}}$};
	\draw[<->] (2.5,1.5) -- (2.5,-1.5) node[midway, above left, yshift=3mm, xshift=1mm] {\footnotesize $(jk)^{\nu_d}\odot$};
	\draw[<-] (6,1.5) -- (6,-1.5) node[midway, above left, yshift=3mm, xshift=1mm] {\footnotesize FFT$^{-1}$};
	\draw[<-] (5.5,-4.5) -- (8.5,-4.5) node[midway, above] {\footnotesize FFT$^{-1}$};
	\node[anchor=north, yshift=-1mm] at (y.south) {$O(N^D \log N)$};
	\node[anchor=north, yshift=-1mm] at (Y.south) {$O(N^D)$};
	\node[anchor=north, yshift=-1mm] at (jkY.south) {$O(N^D \log N)$};
	\node at (0, -4.25) {\parbox{19mm}{repeat for $D$ dimensions}};
\end{tikzpicture}
\end{center}

If we add up the costs, we can see that it's actually no more or less efficient to differentiate along all dimensions at once versus in series.

From a user-friendliness perspective, I judge it to be somewhat more confusing to specify multiple derivative dimensions at once (although generalizing the \texttt{order} and \texttt{axis} parameters to vectors is possible), so I have chosen to limit the package to differentation along a single dimension at a time, which also agrees with the interface of \texttt{chebder}. 

Multidimensional \textit{data} can still be handled, however, via clever indexing and use of \texttt{fft}, \texttt{dct}, and \texttt{chebder}'s \texttt{axis} parameter.

\section{Arbitrary Domains}

So far we've only used the domain $[0, 2\pi)$ in \hyperref[dft]{the Fourier case}, because this is the domain assumed by the DFT, and the domain $[-1, 1]$ in \hyperref[cylinder]{the Chebyshev case}, because this is the where a cosine wrapped around a cylinder casts a shadow. As you may have guessed, this hasn't curtailed the generality of the methods at all, because we can \textit{map} any domain from $a$ to $b$ onto a canonical domain.

\subsection{Fourier on [a, b)}

Say we have $t \in [a, b)$ that we need to map to $\theta \in [0, 2\pi)$. We can accomplish this with:

$$\theta \in [0, 2\pi) \leftrightarrow t \in [a, b) = \underbrace{[0, 2\pi)}_{\theta} \cdot \frac{b - a}{2\pi} + a$$

\noindent To get a sense this is true, let's see an example:

\begin{figure}[h!]
\begin{center}
\begin{minipage}{0.49\textwidth}
\centering
\begin{tikzpicture}
	\begin{axis}[axis lines=middle, xlabel={$t$}, xlabel style={right},
		xmin=0, xmax=8,
		ymin=-4, ymax=4,
		grid=both,
		width=8cm,
		height=5cm]
		\addplot[domain=4:8, samples=100, thick] {cos(deg(pi/2*x + pi/6)) + 3*sin(deg(3*pi/2*x + pi/4))};
	\end{axis}
	\node at (3,3.75) {$\cos(\frac{\pi}{2}t + \frac{\pi}{6}) + 2\sin(\frac{3\pi}{2}t + \frac{\pi}{4}),\ t \in [4, 8)$};
\end{tikzpicture}
\end{minipage}
\begin{minipage}{0.49\textwidth}
\centering
\begin{tikzpicture}
	\begin{axis}[axis lines=middle, xlabel={$\theta$}, xlabel style={right},
		xmin=0, xmax=8,
		ymin=-4, ymax=4,
		grid=both,
		width=8cm,
		height=5cm]
		\addplot[domain=0:2*pi, samples=100, thick] {cos(deg(x + pi/6)) + 3*sin(deg(3*x + pi/4))};
	\end{axis}
	\node at (3.25,3.75) {$\cos(\theta + \cancel{2\pi} + \frac{\pi}{6}) + 2\sin(3\theta + \cancel{6\pi} + \frac{\pi}{4}),\ \theta \in [0, 2\pi)$};
\end{tikzpicture}
\end{minipage}
\begin{tikzpicture}[overlay, remember picture]
	\draw[->, thick] (-10.2, 1.9) -- (-8.4, 1.9) node[midway, above] {$t = \theta \frac{8-4}{2\pi} + 4$};
	\node at (-5.3, 2.7) {\parbox{20mm}{\footnotesize \centering phase is same modulo $2\pi$}};
	\draw[->] (-6.1, 2.5) -- (-6.6, 2.2);
	\draw[->] (-4.5, 2.5) -- (-3.8, 2.2);
\end{tikzpicture}
\end{center}
\end{figure}

\noindent In the discrete case, where we have $M$ samples on $[a, b)$, then we can map $t_n$ with:

$$\theta_n \in \frac{\{0, ... M-1\} \cdot 2\pi}{M} \leftrightarrow t_n \in \frac{\{0, ... M-1\} \cdot \cancel{2\pi}}{M} \cdot \frac{b - a}{\cancel{2\pi}} + a$$

In simple code terms, if we want to take a spectral derivative of a function that's periodic on $[a, b)$, then we need to sample it at \texttt{t\char`_n = np.arange(M)/M * (b - a) + a = np.linspace(a, b, M, endpoint=False)}.

\subsection{Chebyshev on [a, b]}

Here both ends are \textit{inclusive}, so we have $t \in [a, b]$ that we need to map to $x \in [-1, 1]$. We can accomplish this with:

$$x \in [-1, 1] \leftrightarrow t \in [a,b] = \underbrace{[-1, 1]}_{x} \cdot \frac{b - a}{2} + \frac{b + a}{2}$$

\noindent An example to demonstrate this is given in the next figure.

\begin{figure}[h!]
\begin{center}
\begin{minipage}{0.49\textwidth}
\centering
\begin{tikzpicture}
	\begin{axis}[axis lines=middle, xlabel={$t$}, xlabel style={right},
		xmin=-1, xmax=4,
		ymin=-35, ymax=55,
		grid=both,
		width=8cm,
		height=5cm]
		\addplot[domain=1:4, samples=100, thick] {e^x*sin(deg(5*x))};
	\end{axis}
	\node at (3.75,3.75) {$e^t\sin(5t),\ t \in [1, 4)$};
\end{tikzpicture}
\end{minipage}
\begin{minipage}{0.49\textwidth}
\centering
\begin{tikzpicture}
	\begin{axis}[axis lines=middle, xlabel={$x$}, xlabel style={right},
		xmin=-1, xmax=4,
		ymin=-35, ymax=55,
		grid=both,
		width=8cm,
		height=5cm]
		\addplot[domain=-1:1, samples=100, thick] {e^(3/2*x + 5/2)*sin(deg(5*(3/2*x + 5/2)))};
	\end{axis}
	\node at (3.25,3.75) {$e^{(\frac{3}{2}x + \frac{5}{2})}\sin(5(\frac{3}{2}x+\frac{5}{2})),\ x \in [-1,1]$};
\end{tikzpicture}
\end{minipage}
\begin{tikzpicture}[overlay, remember picture]
	\draw[->, thick] (-11, 1.8) -- (-7.6, 1.8) node[midway, above] {$t = x \frac{4-1}{2} + \frac{4+1}{2}$};
\end{tikzpicture}
\end{center}
\end{figure}

In the discrete case, where we have $N+1$ samples on $[a, b]$, then we can map $t_n$ with:

$$x_n \in \cos\!\Big(\frac{\pi \{0, ... N\}}{N}\Big) \leftrightarrow t_n \in \cos\!\Big(\frac{\pi \{0, ... N\}}{N}\Big) \cdot \frac{b - a}{2} + \frac{b + a}{2}$$

\noindent In code this is \texttt{t\char`_n = np.cos(np.arange(N+1)*np.pi/N) * (b - a)/2 + (b + a)/2}.

Notice the order has flipped here, that counting \textit{up} in $n$ means we traverse $x$ from $+1 \rightarrow -1$. This is actually what we want; it corresponds to the \hyperref[cylinder]{horizontal flip} necessary to make cosine shadows equate with Chebyshev polynomails.

\subsection{Accounting for Smoosh}

When a function is sampled at one of the $t_n$ above, then it is \textit{as if} the function lives on the canonical domain. The actual mapping is purely notional, and the spectral differentation procedure proceeds completely agnostic to where the data really came from.

This means the result will actually be the derivative of the smooshed or stretched version of the function on the canonical domain. As the examples hopefully clarified, the \textit{height} of this smooshed function is exactly as it was before, but the \textit{width} is compressed or expanded by a factor of:

$$\text{smoosh} = \frac{\text{length of new interval}}{\text{length of old interval}} = \begin{cases} \frac{2\pi}{b-a} & \text{for Fourier} \\ \frac{2}{b-a} & \text{for Chebyshev} \end{cases}$$

Because a derivative is calculating slope, and slope is rise over run, the answer is effectively now

$$\frac{dy}{dx \cdot \frac{2 \text{ or } 2\pi}{b-a}} = \frac{dy}{dx} \cdot \underbrace{\frac{b-a}{2 \text{ or } 2\pi}}_{\text{scale}}$$

In other words, the overall derivative is scaled by the inverse of the width-smoosh. So to recover the true derivative we want, $\frac{dy}{dx}$, we have to divide by this scale, which is a familiar term from our variable transformations $t \leftrightarrow \theta \text{ or } x$.

For higher derivatives:

$$\frac{d^\nu y}{\big( dx \cdot \text{smoosh})^\nu} = \frac{d^\nu y}{dx^\nu} \cdot \text{scale}^\nu$$

\noindent So we can always correct the derivative by dividing by scale$^\nu$.

To enable calculation of the \texttt{scale}, and to double check the user sampled their function at a correct \texttt{t\char`_n} (especially in the Chebyshev case, since cosine-spacing is easy to flub and especially confusing with the DCT-II), the functions take the sample locations \href{https://pavelkomarov.com/spectral-derivatives/specderiv.html}{as a parameter} and raise error messages with correct examples if the sampling is invalid.

\section{Differentiating in the Presence of Noise}

Finding the true derivative of a noisy signal is ill-posed, because random variations cause unknown (and often dramamtic) local deviations of function slope. This problem only gets worse for calculations of curvature and higher order derivatives, so the only solution is to try to remove the noise \textit{before} differentating.

\subsection{White Noise}

Absent further information about a noise-generating process, we can't assume anything about the noise's structure, which makes it extra challenging to remove. We instead can only rely on the aid of a central observation from the field of signal processing:

\begin{quotation}
``Every spectrum of real noise falls off reasonably rapidly as you go to infinite frequencies, or else it would have infinite energy. But the sampling process aliases higher frequencies in lower ones, and the folding ... tends to produce a flat spectrum. ... \textit{white noise}. The signal, usually, is mainly in the lower frequencies." --Richard Hamming, \textit{The Art of Doing Science and Engineering}\cite{hamming}, Digital Filters III
\end{quotation}

By use of the term ``frequencies", Hamming is implying use of the \textit{Fourier} basis. He's saying there is \textit{band-separation} of signal and noise in the frequency domain, yet another reason for its \hyperref[whyfourier]{popularity}. This general principle extends to other noise reduction techniques in spirit: At bottom, all accomplish some kind of \textit{smoothing}, be it moving average, FIR filtering, Savitzky-Golay, Kalman filtering, total variation penalty, etc.

\subsection{Filtering with the Fourier Basis}

It's helpful to see a picture:

\begin{figure}[h!]\label{spectrum}
\begin{center}
\begin{minipage}{0.6\textwidth}
\centering
\begin{tikzpicture}
  \begin{axis}[width=10cm, height=5cm,
    xlabel={Frequency $f$ (Hz)}, ylabel={Energy}, ylabel style={yshift=-9mm},
    xmin=0, xmax=1.5, xtick={0, 0.5, 1, 1.5}, xticklabels={$0$, $\frac{1}{2}f_s$, $f_s$, $\frac{3}{2}f_s$},
    ymin=0, ymax=1.1, yticklabels={},
    domain=0:1.5,
    samples=200,thick,
    grid=both, axis lines=left]
  \addplot[blue, thick] coordinates {(0.00, 0.8) (0.0625, 1) (0.125, 0.95) (0.1875, 0.8) (0.25, 0.25)
  	(0.3125, 0.27) (0.375, 0.22) (0.4375, 0.2) (0.5, 0.18) (0.5625, 0.21) (0.625, 0.17) (0.6875, 0.18)
  	(0.75, 0.15) (0.8125, 0.11) (0.875, 0.12) (0.9375, 0.15) (1, 0.1) (1.0625, 0.09) (1.125, 0.07)
  	(1.1875, 0.08) (1.25, 0.06) (1.3125, 0.06) (1.375, 0.04) (1.4375, 0.07) (1.5, 0.03)};
  \addplot[dashed, black] coordinates {(0.5,0) (0.5,1.1)};
  \node[black, right] at (axis cs:0.5,0.75) {Cutoff Frequency};
  \end{axis}
\end{tikzpicture}
\end{minipage}
\begin{minipage}{0.35\textwidth}
\centering
\begin{tikzpicture}
  \begin{axis}[width=6cm, height=5cm,
    xlabel={Frequency $f$ (Hz)},
    xmin=0, xmax=0.5, xtick={0, 0.25, 0.5}, xticklabels={$0$, $\frac{1}{4}f_s$, $\frac{1}{2}f_s$},
    ymin=0, ymax=1.1, yticklabels={},
    domain=0:0.5,
    samples=200,thick,
    grid=both, axis lines=left]
  \addplot[blue, thick] coordinates {(0, 0.8) (0.0625, 1) (0.125, 0.95) (0.1875, 0.8) (0.25, 0.25)
  	(0.3125, 0.27) (0.375, 0.22) (0.4375, 0.2) (0.5, 0.18) (0.4375, 0.21) (0.375, 0.17) (0.3125, 0.18)
  	(0.25, 0.15) (0.1875, 0.11) (0.125, 0.12) (0.0625, 0.15) (0, 0.1) (0.0625, 0.09) (0.125, 0.07)
  	(0.1875, 0.08) (0.25, 0.06) (0.3125, 0.06) (0.375, 0.04) (0.4375, 0.07) (0.5, 0.03)};
  \addplot[dashed, black] coordinates {(0.5,0) (0.5,1.1)};
  \node[black, right] at (axis cs:0.05,0.25) {Aliasing};
  \end{axis}
\end{tikzpicture}
\end{minipage}
\begin{tikzpicture}[overlay, remember picture]
	\draw[->, thick] (-7.3, 0.5) -- (-6, 0.5);
\end{tikzpicture}
\end{center}
\vspace{-5mm}
\captionsetup{width=0.9\textwidth}
\caption*{\small Typical energy spectrum of a noisy signal, before and after sampling. $f_s$ is a sampling rate of our choosing.}
\end{figure}

In practice we expect to sample a signal frequently enough to capture fluctuations of interest in the data, so signal energy should be concentrated in frequencies below some cutoff.  Noise energy, decaying up to higher frequencies, is added in linearly.

To reach the frequency domain, we use the FFT, which requires \textit{equispaced} samples and a \textit{periodic} signal, because a discontinuity or corner causes \hyperref[artefacts]{artefacts at higher frequency}, which become impossible to model in finitely many coefficients and impossible to distinguish from noise. Nyquist theorem\cite{oppenheim} tells us that we need $>\!2$ equispaced samples per cycle to unambiguously reconstruct a frequency, so this process creates a natural cutoff at the Nyquist rate, $f_s/2$. Equispaced samples taken from frequencies slightly over this limit are \hyperref[bandlimited]{best matched} by frequencies slightly under the limit, just as unitary complex numbers alias: $e^{j(\pi + \epsilon)} = e^{-j(\pi - \epsilon)}$. This creates a \textit{folding} pattern in the bandlimited FFT spectrum, which distributes decaying noise energy somewhat evenly.

Notice the more we sample, the more we can concentrate the legitimate signal's energy in low frequencies, and the more we can distribute noise energy across higher frequencies. We can then \textit{zero out} or otherwise dampen the upper part of the spectrum to chop down the noise that hasn't aliased on to the periodic signal's band! The \href{http://pavelkomarov.com/spectral-derivatives/specderiv.html}{\texttt{filter} parameter} is designed exactly for this. If we then inverse transform, we get a smoother signal, or we can multiply by $(jk)^\nu$ and then inverse transform to sample the derivative of that smoother signal.

\subsection{The Advantage of Spectral Smoothing}

Most alternative noise-quelling techniques can only take advantage of local information, but a spectral representation builds a function out of basis functions that span the entire domain, so every point's value takes holistic fit under consideration. This makes the reconstruction much more \textit{robust} to perturbations than one that uses only a few neighboring points. I.e.~it's much harder to \textit{corrupt} the signal so thoroughly that it can't be successfully recovered.

In the Fourier case, this has an analogy with \href{https://www.youtube.com/watch?v=X8jsijhllIA}{error-correcting codes}\cite{sanderson, hamming}; notice the locations corresponding to Hamming code parity checks constitute selections of different frequency:

\begin{center}
\begin{tikzpicture}[scale=0.6]
	\foreach \x in {1,3} { 
		\fill[gray!50] (\x,0) rectangle (\x+1,4);}
	\fill[gray!50] (7,0) rectangle (9,4);
	\foreach \y in {0,2} {
		\fill[gray!50] (10,\y) rectangle (14,\y+1);}
	\fill[gray!50] (15,0) rectangle (19,2);
	\foreach \xy in {0,1,2,3,4} { 
		\foreach \offset in {0,5,10,15} { 
			\draw (\xy+\offset,0) -- (\xy+\offset,4);
			\draw (0+\offset,\xy) -- (4+\offset,\xy);}}
	\foreach \offset in {0,5,10,15} {
		\foreach \i in {0,1,2,3} {
			\foreach \j in {0,1,2,3} {
				\pgfmathtruncatemacro{\n}{\i*4+\j+1}
				\node at (\j+0.75+\offset,3.75-\i) {\footnotesize \n};}}}
	\node at (2, -0.5) {check \#1};
	\node at (7, -0.5) {check \#2};
	\node at (12, -0.5) {check \#3};
	\node at (17, -0.5) {check \#4};
\end{tikzpicture}
\vspace{-2mm}
\end{center}

Of course there are distinctions: In the context of continuous signals, ``corruption" means (discrete representations of) continous numbers have slightly inaccurate values, whereas in error correcting codes each datum is a single discrete bit which can simply be flipped. Likewise, a Fourier coefficient indicates ``We need this much of the $k^\text{th}$ basis function", whereas the analogous parity check indicates ``There is/isn't a parity error among my bits."

But in both cases a spot-corruption will stand out, because it appears in a particular combination of parity checks or introduces a value that can't as easily be represented with a finite combination of smooth basis functions. In this sense, the Fourier basis is \textit{the ideal} basis for filtering high frequency noise, so long as the signal of interest is periodic, and other methods are merely trying to approximate a lowpass filter.

\subsection{Measurement versus Process Uncertainty}

Noise can have a couple different sources. If noise is due to imperfect sensing, then each sample is drawn from a random variable. By contrast, if there is uncertainty in the underlying dynamics, then we might say there is noise in the process itself. If we sample our signal uniformly, as we do when using the Fourier basis and most other filtering methods, since they tend to be more computationally expensive to generalize to non-uniform samplings\cite{nonuniform}, then we don't really need to make a distinction between these noise sources, because they look the same. But if we cosine-sample our data for expediency when using the Chebyshev basis, there will be a difference, because samples taken closer together will have greater correlation.

\begin{center}
\begin{minipage}{0.32\textwidth}
\centering
\begin{tikzpicture}
  \begin{axis}[width=7cm, height=4cm,
    xlabel={\small \shortstack{cosine-spaced noise\\measurement noise in $x$ domain\\\phantom{0}}}, xlabel style={yshift=5mm},
    xmin=0, xmax=101, xticklabels={},
    ymin=-2.5, ymax=3, yticklabels={}]
 	\addplot[blue, thick] coordinates {
    (1, 0.44924) (1.02467, -0.99617) (1.09866, 1.57128) (1.22190, 0.52507) (1.39426, -1.28970)
    (1.61558, -1.44505) (1.88564, 0.27104) (2.20416, 1.43915) (2.57084, -0.13549) (2.98532, -0.09529)
    (3.44717, -1.07594) (3.95596, 1.26427) (4.51118, 0.07829) (5.11227, 1.00133) (5.75865, -0.69820)
    (6.44967, 0.09316) (7.18467, -0.47154) (7.96290, -0.45974) (8.78360, 1.46259) (9.64597, -0.72314)
    (10.54915, -0.62851) (11.49225, -0.55570) (12.47434, 0.02975) (13.49445, -0.05741) (14.55157, 0.12502)
    (15.64466, 1.32299) (16.77264, -0.11373) (17.93441, 0.47265) (19.12880, 0.13899) (20.35465, 0.55774)
    (21.61074, -1.31526) (22.89583, 0.69607) (24.20866, -0.28113) (25.54793, 0.39325) (26.91232, 1.00132)
    (28.30048, 0.27553) (29.71104, -0.68995) (31.14261, -0.41917) (32.59377, -0.66658) (34.06310, -0.39784)
    (35.54915, 0.07289) (37.05044, 1.65013) (38.56551, 1.58383) (40.09284, -1.60049) (41.63093, -0.15252)
    (43.17828, 0.29528) (44.73334, 0.70752) (46.29458, -0.36491) (47.86047, -0.60486) (49.42946, -1.99844)
    (51, 1.21089) (52.57054, 0.80667) (54.13953, 0.79436) (55.70542, 0.36514) (57.26666, 0.87209)
    (58.82172, -1.30100) (60.36907, 0.63087) (61.90716, 0.72403) (63.43449, -0.44550) (64.94956, 1.28720)
    (66.45085, 0.70524) (67.93690, -1.32819) (69.40623, -1.39457) (70.85739, -1.30535) (72.28896, -2.20101)
    (73.69952, -0.41779) (75.08768, 1.51886) (76.45207, 0.06770) (77.79134, 1.36965) (79.10417, -0.13765)
    (80.38926, 0.21485) (81.64535, -0.67176) (82.87120, 1.17936) (84.06559, -0.01664) (85.22736, -0.79795)
    (86.35534, -0.86843) (87.44843, -1.47980) (88.50555, 1.12616) (89.52566, -0.44343) (90.50775, 0.55058)
    (91.45085, -0.39320) (92.35403, -0.96446) (93.21640, 0.47854) (94.03710, -0.34660) (94.81533, -0.09097)
    (95.55033, -0.24364) (96.24135, 0.36869) (96.88773, 1.27397) (97.48882, -0.08068) (98.04404, 1.16712)
    (98.55283, -0.85430) (99.01468, -1.33674) (99.42916, 2.06015) (99.79584, 0.19364) (100.11436, 1.02679)
    (100.38442, -0.80207) (100.60574, 0.80308) (100.77810, 3.02958) (100.90134, -0.50121) (100.97533, -1.25102)
    (101, 0.83996)};
  \end{axis}
\end{tikzpicture}
\end{minipage}
\begin{minipage}{0.32\textwidth}
\centering
\begin{tikzpicture}
  \begin{axis}[width=7cm, height=4cm,
    xlabel={\small \shortstack{equispaced noise\\measurement noise in $\theta$ domain or\\process noise in $x$ domain}},
    xlabel style={yshift=5mm},
    xmin=0, xmax=101, xticklabels={},
    ymin=-2.5, ymax=3, yticklabels={}]
 	\addplot[blue, thick] coordinates {
    (1, 0.44924) (2, -0.99617) (3, 1.57128) (4, 0.52507) (5, -1.28970)
    (6, -1.44505) (7, 0.27104) (8, 1.43915) (9, -0.13549) (10, -0.09529)
    (11, -1.07594) (12, 1.26427) (13, 0.07829) (14, 1.00133) (15, -0.69820)
    (16, 0.09316) (17, -0.47154) (18, -0.45974) (19, 1.46259) (20, -0.72314)
    (21, -0.62851) (22, -0.55570) (23, 0.02975) (24, -0.05741) (25, 0.12502)
    (26, 1.32299) (27, -0.11373) (28, 0.47265) (29, 0.13899) (30, 0.55774)
    (31, -1.31526) (32, 0.69607) (33, -0.28113) (34, 0.39325) (35, 1.00132)
    (36, 0.27553) (37, -0.68995) (38, -0.41917) (39, -0.66658) (40, -0.39784)
    (41, 0.07289) (42, 1.65013) (43, 1.58383) (44, -1.60049) (45, -0.15252)
    (46, 0.29528) (47, 0.70752) (48, -0.36491) (49, -0.60486) (50, -1.99844)
    (51, 1.21089) (52, 0.80667) (53, 0.79436) (54, 0.36514) (55, 0.87209)
    (56, -1.30100) (57, 0.63087) (58, 0.72403) (59, -0.44550) (60, 1.28720)
    (61, 0.70524) (62, -1.32819) (63, -1.39457) (64, -1.30535) (65, -2.20101)
    (66, -0.41779) (67, 1.51886) (68, 0.06770) (69, 1.36965) (70, -0.13765)
    (71, 0.21485) (72, -0.67176) (73, 1.17936) (74, -0.01664) (75, -0.79795)
    (76, -0.86843) (77, -1.47980) (78, 1.12616) (79, -0.44343) (80, 0.55058)
    (81, -0.39320) (82, -0.96446) (83, 0.47854) (84, -0.34660) (85, -0.09097)
    (86, -0.24364) (87, 0.36869) (88, 1.27397) (89, -0.08068) (90, 1.16712)
    (91, -0.85430) (92, -1.33674) (93, 2.06015) (94, 0.19364) (95, 1.02679)
    (96, -0.80207) (97, 0.80308) (98, 3.02958) (99, -0.50121) (100, -1.25102)
    (101, 0.83996)};
  \end{axis}
\end{tikzpicture}
\end{minipage}
\begin{minipage}{0.32\textwidth}
\centering
\begin{tikzpicture}
  \begin{axis}[width=7cm, height=4cm,
    xlabel={\small \shortstack{cosine-sampling of equispaced noise\\process noise in $\theta$ domain\\\phantom{0}}},
    xlabel style={yshift=5mm},
    xmin=0, xmax=101, xticklabels={},
    ymin=-2.5, ymax=3, yticklabels={}]
 	\addplot[blue, thick] coordinates {
    (1, 0.83996) (2, 0.78837) (3, 0.63365) (4, 0.37597) (5, 0.01556)
    (6, -0.44721) (7, -1.01189) (8, -1.09794) (9, -0.82300) (10, -0.51222)
    (11, 1.07767) (12, 2.87409) (13, 1.89145) (14, 0.62287) (15, -0.41466)
    (16, 0.02032) (17, 0.87294) (18, 0.22456) (19, 1.65625) (20, -0.13414)
    (21, -1.07181) (22, 0.14074) (23, 0.57524) (24, 0.58912) (25, 0.77465)
    (26, -0.02606) (27, -0.12568) (28, -0.32984) (29, 0.29268) (30, -0.76186)
    (31, 0.18320) (32, -0.33988) (33, 0.58240) (34, -1.14481) (35, -0.80413)
    (36, 0.34273) (37, -0.13685) (38, 0.16458) (39, 0.75734) (40, 0.15927)
    (41, 0.45535) (42, -2.15583) (43, -1.35580) (44, -1.13941) (45, 1.07242)
    (46, -0.23700) (47, 0.65571) (48, -0.66084) (49, 0.43587) (50, 0.79965)
    (51, 1.21089) (52, -1.20335) (53, -0.21528) (54, 0.41672) (55, -0.53864)
    (56, 1.01614) (57, 1.06803) (58, -0.35413) (59, -0.55908) (60, -0.67629)
    (61, 0.60275) (62, 0.43162) (63, 0.11584) (64, -1.02843) (65, 0.43674)
    (66, 0.37239) (67, 0.01225) (68, 0.78143) (69, -0.01935) (70, -0.03124)
    (71, -0.58404) (72, -0.68958) (73, 1.18107) (74, -0.46051) (75, -0.34315)
    (76, -0.18804) (77, 0.06392) (78, 0.53468) (79, 0.70172) (80, 0.07603)
    (81, -0.63381) (82, -0.10952) (83, 0.20526) (84, 1.39582) (85, 0.48675)
    (86, -0.67337) (87, -1.40756) (88, -1.30714) (89, -0.40260) (90, 0.57114)
    (91, 1.10344) (92, 1.53358) (93, 0.46944) (94, -0.47199) (95, -0.83087)
    (96, -0.44053) (97, -0.12063) (98, 0.12850) (99, 0.30663) (100, 0.41358)
    (101, 0.44924)};
  \end{axis}
\end{tikzpicture}
\end{minipage}
\end{center}

When we sample at cosine-space points, $\Delta x$ decreases near the edges. In the case of measurement noise, effects on $\Delta y$ are undiminished in these regions, so the error in $\frac{\Delta y}{\Delta x}$ increases dramatically. In the case of process noise the situation is a little more hopeful, because its effects on $\Delta y$ naturally get smaller as $\Delta x$ shrinks.

\subsection{Filtering with Polynomial Bases (is Terrible)}

To avert Gibbs phenomenon in the case of aperiodic data or to achieve better compression (more energy represented in fewer modes) in case we know something about how our data should look, we may prefer to use another basis, like Chebyshev polynomials or PCA modes for representation. All bases have ``lower frequency" and ``higher frequency" elements, meaning some modes have fewer transitions between low and high values, and some have more. Similar to Fourier basis representations, signal energy empirically tends to cluster in lower modes, and noise tends to be scattered across modes, similar to the \hyperref[spectrum]{typical spectrum} above.

However, smoothing aperiodic data is fundamentally more difficult than the periodic case at the edges of the domain, because in the periodic case we can smooth across the domain boundary, whereas with an aperiodic function we only get one-sided information about those regions. And unlike Fourier basis functions, which have uniform frequency throughout their domain, other bases are typically nonuniform. This may be desirable if the data or noise is known to fit a particular pattern, but in the naive case, where we don't know the noise shape or expect white noise, nonuniform basis functions' variable expressive power across the domain can mismatch the situation. Chebyshev polynomials exhibit this characteristic, getting steeper and therefore higher frequency near the boundaries, which makes them worse at filtering and better at \textit{fitting} high frequency noise in these regions, i.e.~more \textit{sensitive} to disruptions there.

\begin{figure}[h!]
	\centering
	\includegraphics[width=\textwidth]{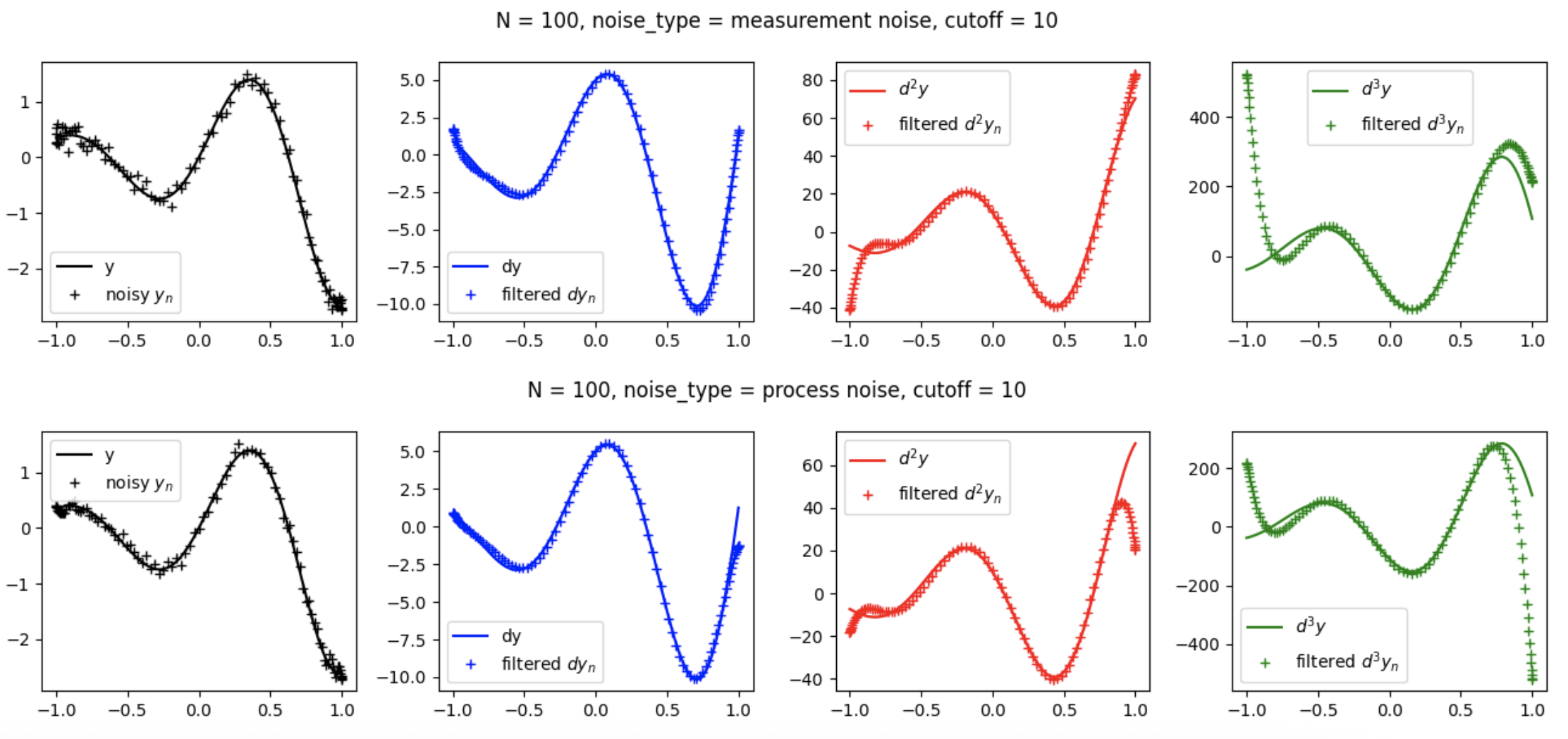}
	\caption*{\small An example of systematic edge blowup for Chebyshev derivatives in the presence of noise.}
\end{figure}

This problem is perhaps best shown mathematically in the \hyperref[algo]{Chebyshev-via-Fourier method}, where we not only implicitly warp the function to be $y(\cos(\theta))$ by taking a DCT (thereby treating measurement noise as if it's evenly distributed in the $\theta$ domain), but we later also explicitly \textit{unwarp} to get back to the $x$ domain, involving \hyperref[pyramid]{division by powers of $\sqrt{1-x^2}$}\cite{brown, trefethen8}, which $ \rightarrow 0$ as $x \rightarrow \pm 1$. Numerically equivalent, in the \hyperref[seriesrule]{Chebyshev series-based method}, ``Tight coupling between coefficients enables propagation of errors from high frequency to low frequency modes."\cite{brown} We see that higher-order coefficients have an impact on \href{https://scicomp.stackexchange.com/q/44939/48402}{ever-other lower-order coefficient}\cite{dcoefs}, and this effect is cumulative, so a slight error in coefficient values, especially higher-order ones, compounds in a very nasty way when we differentiate. Thus ``while the representation of a function by a Chebyshev series may be most accurate near $x = \pm 1$, these results indicate that the derivatives computed from a Chebyshev series are \textit{least accurate} at the edges of the domain."\cite{brown}

In fact, the Chebyshev basis is not the only one suffer this weakness, because other polynomial bases, e.g. the Legendre polynomials (orthogonal without a weighting function) and \href{https://en.wikipedia.org/wiki/Bernstein_polynomial}{Bernstein polynomials} (linearly independent but not orthogonal), experimented with in the \href{https://github.com/pavelkomarov/spectral-derivatives/blob/main/notebooks/filtering_noise.ipynb}{filtering noise notebook}, are also more sensitive at the domain edges, even when sampled at cosine-spaced points to help dampen Runge phenomenon. This is a fundamental limitation of polynomial-based methods in the presence of noise. Occasionally differentiation can be made to work okay by filtering higher modes, but there is always \href{https://github.com/pavelkomarov/spectral-derivatives/blob/main/notebooks/filtering_noise.ipynb}{systematic blowup} in higher-order derivatives.

\printendnotes

\end{document}